\documentclass{elsarticle}

\usepackage[utf8]{inputenc}
\usepackage{ulem}
\usepackage{color}

\usepackage{slashed}

\usepackage[retainorgcmds]{IEEEtrantools}	

\usepackage{amsmath, amsthm, amsfonts,amssymb}

\journal{Nuclear Physics B}

\begin{document}

  \begin{frontmatter}
  
    \title{Lagrangians for Massive Dirac Chiral Superfields}

   \author[a]{Enrique~Jim\'enez}      
     \address[a]{Instituto de F\'isica, Universidad Nacional Aut\'onoma de M\'exico, Apdo. Postal 20-364, 01000, M\'exico D.F., MEXICO}
   \ead{ejimenez@fisica.unam.mx}

   \author[b,c]{C.A.~Vaquera-Araujo}
   \address[b]{ Facultad de Ciencias, CUICBAS, Universidad de Colima, 28040 Colima, MEXICO}
\address[c]{AHEP Group, Institut de F\'{i}sica Corpuscular -- C.S.I.C./Universitat de Val\`{e}ncia, \\ Parc Cientific de Paterna.   C/Catedratico Jos\'e Beltr\'an, 2 E-46980 Paterna (Val\`{e}ncia) - SPAIN}
   \ead{vaquera@ific.uv.es}

   \begin{abstract}
{A variant for the superspin one-half massive  superparticle in  $ 4D $, $ \mathcal{N}=1 $, based on  Dirac superfields,  is offered. As opposed to the current known models that use spinor chiral superfields, the propagating fields of the supermultiplet are those of the lowest mass dimensions possible: scalar, Dirac and vector fields. Besides the supersymmetric chiral condition, the Dirac superfields are not further constrained,  allowing a very straightforward implementation of the path-integral method. The corresponding superpropagators are  presented. In addition, an interaction  super Yukawa potential, formed by   Dirac and scalar chiral superfields, is given in terms of their component fields. The model is first presented for the case of two superspin one-half superparticles related by the charged conjugation operator, but in order to treat the case of neutral superparticles, the Majorana condition on the Dirac superfields is also studied.  We compare our proposal with the known models of spinor superfields for the one-half superparticle and show that it is equivalent to them.
}
\end{abstract}
\end{frontmatter}


\section{Introduction}
\label{sec:intro}
More than forty years after  its invention, supersymmetry still possesses some unexplored and/or not completely understood facets, providing a source of active research such as the study of massive supersymmetric theories. To gain further insight on the subject, we present a new framework for superspin one-half in $ 4D $, $ \mathcal{N}=1 $ (see~\cite{Wess:1992cp, Buchbinder:1995uq} for examples of well known free superspin one-half models).  Over the  last decade progress has been achieved  in the formulation of free massive theories up to superspin   three-halves \cite{Ogievetsky:1976qb,Altendorfer:1999mn,Buchbinder:2002gh,Buchbinder:2002tt,Gregoire:2004ic,Buchbinder:2005je,Gates:2005su,Gates:2006cq,Gates:2013tka}. Generalizations to the case of complex mass can be found in \cite{D'Auria:2004sy,Louis:2004xi,Theis:2004pa,Kuzenko:2004tn}.  A common feature of these studies is the  use  of general   off-shell superfields  restricted only by their reality condition.  Their convenience relies on the fact that taking appropriate products of superderivatives on these superfields, one can create constrained (by construction) spinor chiral superfields that  have a smooth zero mass limit. 
An  example of this situation is the relation between left and right  strength superfields (Bianchi identities) in  supersymmetric gauge theories.   

 We approach the study  of massive one-half supermultiplets by working with   spinor chiral superfields that are not further constrained. There are two  models for these supermultiplets based on chiral spinor superfields \cite{Siegel:1979ai,Gates:1980az}. A distinctive feature
of our model is that auxiliary fermionic fields are essential for the
off-shell closure of the superalgebra, and they play a key role in providing the mass terms of the propagating fermionic degrees of freedom, a situation that is not present in the current known models. The degrees of freedom of the component fields for the one-half superspin models in \cite{Siegel:1979ai,Gates:1980az},  represent a 8(fermionic)+8(bosonic) realization of supersymmetry, the present work  represents a new off-shell  (16+16) irreducible realization of supersymmetry for the superspin one-half case.

Differences between models describing the same supermultiplet rely on the possible couplings of the corresponding interacting theory, and those couplings are in turn determined by the propagating fields that carry the particles of the supermultiplet.  In~\cite{Siegel:1979ai},  the spin-one particles propagate trough second-rank antisymmetric tensors.  In addition, the spin-zero state in \cite{Gates:1980az} is described by a 3-form field.  As opposed  to these models,  where  second-order derivatives and superderivatives are present in the free action, here we  introduce only first order derivatives (in a similar fashion to Dirac theory in ordinary space), allowing us to represent the  supermultiplet as a collection of fields with the lowest dimension possible: scalar, Dirac and vector fields.

The use of chiral superfields that are not further constrained makes very easy the implementation of the path-integral; the calculation of the superpropagator for the Dirac superfields is carried out on the same lines  of the Wess-Zumino model, opening the possibility of further studies on the renormalization properties of  interacting theories constructed with these superfields. We give the details of  a super Yukawa model, the simplest possible  interacting theory.

Recently, in the context of a superspace formulation of Weinberg's ``noncanonical’’ methods \cite{Weinberg:1969di}, a set of super Feynman rules for arbitrary superspin massive theories has been presented in \cite{Jimenez:2014gfa}, together with the explicit form of the interaction picture
superfields for arbitrary superspin. 
 These superfields have the common feature of being exclusively constrained by the supersymmetric chiral condition, and therefore they bear a closer resemblance to the superfields of the Wess-Zumino model than those superfields constructed through superderivatives of general superfields. 
One of the main purposes of this paper is to construct an off-shell model by
canonical and path-integral methods, beyond the superspin zero case, where the properties of \cite{Jimenez:2014gfa} are met. We show that  both  the  spinor superfields of Ref. \cite{Siegel:1979ai}   and the Dirac chiral superfields, in the interaction picture and the superpropagators
of the models, coincide with those of \cite{Jimenez:2014gfa}, establishing a proof of consistency between
both formalisms.

Difficulties in formulating general results for higher superspins and extended supersymmetries are in part due  to the  issue known as the 'SUSY off-shell challenge' \cite{Buchbinder:2002gh,Gates:2001rn}, that  can be enunciated as follows: given the  propagating  off-shell fields of the  corresponding supersymmetric multiplets, find an enlarged set of  fields such  that the superalgebra   closes. Then write the  corresponding invariant Lagrangian for the enlarged set of fields. In order to maintain  the  given supermultiplets, the required    additional fields must not propagate (they {must} be auxiliary). This last requisite is satisfied if  the equations of motion for the auxiliary fields are algebraic equations.

In contrast to the approach followed by  \cite{Dixon:2015cya}, here we start with a superfield formulation, ensuring the closure of the superalgebra. Then the problem is to elucidate if  the chosen superfields can  generate a sensible Lagrangian, such that its free part (after eliminating the auxiliary fields)   is the correct Lagrangian for the propagating fields.   

When tackling higher  superspin theories of  chiral superfields with odd  spinor  indices, the use of bilinear terms in the K\"ahler potential ($ \mathcal{D} $-terms), gives a second order derivative term for the zeroth component of the fermionic expansion,  leading naturally  to a theory with second order derivatives for fermions. Although there are constructions for these kinds of theories, both in space and in superspace \cite{Wunderle:2010yw},  their unitarity is not manifest \cite{Veltman:1997am}.
A solution  to this  conundrum is based on  the observation that an  $ \mathcal{F} $-term,  with the same  dimensionality as the $ \mathcal{D} $-term, can be introduced with the help of the Dirac derivative $ \slashed{\partial} $. Even though we focus on superfields with just one spinor index, the setting can in principle work for  chiral superfields with an arbitrary odd number of spinor  indices, where the Dirac derivative is available.

In this paper we derive the free action for  Dirac superfields  by requiring the following conditions:
\begin{itemize}
\item[(a)] The absence of second order derivatives for fermion fields  off-shell. 
\item[(b)] The emergence (after eliminating the auxiliary fields) of the free action for Dirac, scalar,  and vector fields with degenerate mass.
\end{itemize}

We introduce two different types of chiral superfields in order to distinguish the particle supermultiplet from the antiparticle one.   Both the first  and second order terms in the fermionic variables expansion contain auxiliary fields,  in contrast to the cases for the scalar and the massless strength chiral superfields, where there are auxiliary fields only in one term for each case.
After identifying  the scalar and vector  propagating fields, we write the off-shell tensor-field (that appears in the linear term of the Dirac  superfields) in terms of the  $4\times 4$  matrix covariant basis $ \left(I,\gamma_{5},\gamma_{\mu},\gamma_{\mu}\gamma_{5},\left[\gamma_{\mu},\gamma_{\nu} \right]  \right)  $.
Taking suitable linear combinations of operators of dimension ${5} $ and less, the desired free Lagrangian is obtained.   Next,  we calculate the
corresponding superpropagators following a straightforward generalization of the chiral
scalar superfield case. We also discuss the similarities and differences of the model~\cite{Siegel:1979ai}(denoted as Siegel model in what follows) and ours. We give expressions for the interacting  potential of a super Yukawa model: a scalar superfield coupled to  a bilinear of Dirac superfields,  and show that this Yukawa term cannot be present in Siegel model. Finally,  imposing the Majorana condition on the Dirac superfield, we obtain the case of a neutral superparticle.

 We adopt the notation and conventions of \cite{Weinberg:2000cr}, reviewed briefly in Appendix A, where we  also include formulae for the tensor-spinor fields used throughout the paper. We have chosen a four-spinor notation for supersymmetry, since this allows us to work in a unified manner with the four-spinor notation of the  Dirac superfields. For the sake of completeness, we have included a section where some of our findings are written in the more common dotted-undotted spinor notation.
 
 Summarizing, in this paper we present a variant realization for the model of two massive superspin-
1/2 multiplets, as well as the special case of just one multiplet. It is shown that this latter variant is equivalent to the known models based on chiral spinor superfields. The distinctive feature of our proposal is the introduction of auxiliary spinor component fields that in turn leads to a description of the physical fields in terms of their lowest dimensional representation. The explicit relation between or model and those described in \cite{Siegel:1979ai,Gates:1980az} is currently under investigation and will be presented elsewhere. 
 
The paper is structured as follows. Section \ref{sec:DiracSuperfields} deals with the general properties of Dirac superfields. The details of the tensor-spinor fields are studied in Section \ref{sec:TensorSpinorfield} and the supersymmetric free Lagrangian is constructed in Section \ref{sec:SupLagrangians}. In Section \ref{sec:SupYuk} we give expressions for  the interacting potential of a super Yukawa kind.   Reality conditions on the superfields are presented in Section \ref{sec:RealityCon}. Section \ref{sec:twoComponent} is devoted to show some of our results in a two-spinor notation.  Finally our conclusions are offered in Section \ref{sec:Conclusions}.

\section{Dirac superfields}
\label{sec:DiracSuperfields}

The  superspin  one-half superparticle consists of states with spin $ \left(\frac{1}{2},\frac{1}{2}, 0,1\right)$ (note that super antiparticles can be different from their superparticles). The corresponding propagating fields  are  denoted by
\begin{IEEEeqnarray}{rl}
            \left( \chi^{+}\right)_{\alpha} ,  \quad \left( \chi^{-}\right)_{\alpha} , \quad \phi, \quad a_{\mu} \ .
    \label{QueAlfin}
\end{IEEEeqnarray}
  We insert the Dirac field   $ \left( \chi^{+}\right)_{\alpha} $  in the zeroth order fermionic component of a $ (+) $-chiral  superfield  $ \left( \Psi^{+} \right) _{\alpha}$. Similarly,   $ \left( \chi^{-}\right)_{\alpha}$  is embedded in a $ (-) $-chiral superfield  $ \left( \Psi^{-} \right) _{\alpha}$.  The signs $\left( \pm\right)  $ refer to the chirality conditions
\begin{IEEEeqnarray}{rl}
             \tfrac{1}{2}\left[ \left(I \, \mp\,\gamma_{5} \right)\mathcal{D}\right]_{\beta}  \left( \Psi^{\pm} \right) _{\alpha}  \, = \, 0\ ,
    \label{SUSYchiralCondition}
\end{IEEEeqnarray}
where  $ \mathcal{D}_{\alpha} $ is the four-spinor superderivative. The rest of the component fields are the auxiliary Dirac fields $ \mathcal{G}^{+} $ and $ \mathcal{G}^{-} $, and a tensor-(four)spinor field $ \xi_{\beta\alpha} $ that contains the physical fields  $ \phi $  and $ a_{\mu} $.  Specifically, the superfields are defined by the expansion
\begin{IEEEeqnarray}{rl}
            \left(  {\Psi}^{\pm} \right)_{\alpha} & \, \equiv\, \left( \chi^{\pm} \right) _{\alpha}   \, - \,  \sqrt{2}\left( \vartheta_{\pm}\cdot \gamma_{5}\, \xi\right) _{\alpha}\, + \, \vartheta\cdot  \vartheta_{\pm}\,\left(  \mathcal{G}^{\pm}  \, + \, \slashed{\partial}\chi^{\mp} \right)_{\alpha} \ , 
     \label{DefDiracSuperField}
 \end{IEEEeqnarray}
with $  2\vartheta_{\pm}  \, = \, \left( I\pm \gamma_{5} \right) \vartheta $. The dot product between two 4-spinors $ \vartheta $ and $ \vartheta' $ is defined by the relation $ \vartheta\cdot \vartheta' \, = \,  \vartheta^{\intercal}\epsilon\gamma_{5} \vartheta' $. Notice that the mass dimension of \eqref{DefDiracSuperField} is $\left[  \frac{3}{2}\right]  $.  Having $ \slashed{\partial}\chi^{\mp} $ in the $ \mathcal{F}_{\pm}$-term (given as the coefficient of  $\vartheta\cdot \vartheta_{\pm}  $) is well-defined, since under infinitesimal supersymmetric transformations,  both the zeroth order term in the fermionic expansion of a $ (\pm) $-chiral superfield,   and the second order term in the expansion of  a $ (\mp) $-chiral superfield, transform linearly in $ \vartheta_{\pm} $.   This shift will be essential to obtain a first order derivative theory for fermion fields off-shell; the coefficients accompanying these derivative terms  are chosen for later convenience. Component fields transform under an infinitesimal supersymmetry transformation as
\begin{IEEEeqnarray}{rl}
 \delta \chi^{\pm}  _{\alpha}   &  \, =\,  \sqrt{2}\left(  \tilde{\vartheta}_{\pm}\cdot \gamma_{5}  \,\xi \right) _{\alpha}\ ,\nonumber \\
            \delta  \mathcal{G}^{\pm}  _{\alpha}  &  \, =\, -\sqrt{2}\left( \tilde{\vartheta}_{\mp} \cdot \gamma_{5}\left[ \slashed{\partial} \,\xi  \, + \, \xi \overleftarrow{\slashed{\partial}}^{\intercal}\right] \right) _{\alpha} \ , \nonumber \\  
                \delta\xi_{\alpha\beta} &  \, =\,  \sqrt{2}\sum^{+,-}_{\varepsilon}\left\lbrace \left(\gamma_{5}\slashed{\partial}  \tilde{\vartheta}_{-\varepsilon}\right)_{\alpha}\left( \chi^{\varepsilon}\right)_{\beta}  \, + \,  \left( \gamma_{5} \tilde{\vartheta}_{\varepsilon}\right)_{\alpha} \left( \mathcal{G}^{\varepsilon} \, + \, \slashed{\partial}\chi^{-\varepsilon}\right)_{\beta} \right\rbrace  \ .
\label{ComponentSUSYTransformations}
\end{IEEEeqnarray}
Here $ \tilde{\vartheta} $ stands for the infinitesimal fermionic parameter and $ \left(\cdots\right)  \overleftarrow{\slashed{\partial} }  \, = \, \partial_{\mu}\left(\cdots \right) \gamma^{\mu} $ .
The  Dirac adjoint of the superfield is defined by the relation
   \begin{IEEEeqnarray}{rl}
             \bar{\Psi}^{\pm} \left( x,\vartheta\right)  \, \equiv \,  \left[ {\Psi}^{\mp} \left( x,\vartheta\right)\right] ^{\dagger}\beta \ ,
     \label{DiracAdjoint}
 \end{IEEEeqnarray}
and reads
 \begin{IEEEeqnarray}{rl}         
               \left( \bar{\Psi}^{\pm} \right) _{\alpha}     & \, =\, \left( \bar{\chi}^{\mp}\right)_{\alpha}  \, - \,  \sqrt{2} \left( \vartheta_{\pm}\cdot \epsilon\,\bar{\xi}\right) _{\alpha}\, + \, \vartheta\cdot  \vartheta_{\pm}\, \left(  \bar{\mathcal{G}}^{\mp}   \, - \, \bar{\chi}^{\pm}\overleftarrow{\slashed{\partial} }\right)_{\alpha} \ ,  \label{EstaConvencidaDeQueEsAsi}
 \end{IEEEeqnarray}
where $  \bar{\xi}  \, = \, \beta \xi^{*} \beta $. From the point of view of Poincaré invariance the tensor-spinor field forms a reducible representation. However, from the point of view of supersymmetry, the Dirac superfields are irreducible since they contain 16 fermionic and 16 bosonic complex components (unless further constraints are considered).

 \section{Tensor-spinor fields}
 \label{sec:TensorSpinorfield}
 In order to identify the propagating fields in $  \xi_{\alpha\beta}$, we study this  tensor-spinor field in
the interaction picture. To do so we use  the methods of Weinberg \cite{Weinberg:1969di}, where the   momentum wavefunctions for the annihilation part of  the free field is given by the product $ {2p^{0}}\,u_{\alpha}\left(\textbf{p},\sigma \right) u_{\beta} \left(\textbf{p},\tilde{\sigma} \right)$ of two Dirac spinors (the corresponding product for  the creation part is  $ {2p^{0}}\,v_{\alpha}\left(\textbf{p},\sigma \right) v_{\beta} \left(\textbf{p},\tilde{\sigma} \right)$). As each spinor index carries a spin one-half, the general  particle content for this  field consists of a particle (and its antiparticle) of spin in  the tensorial representation   $ \frac{1}{2}\otimes\frac{1}{2}  $. By construction,   $ \xi $ satisfies  two Dirac  equations:
\begin{IEEEeqnarray}{rl}
              \left(\slashed{\partial}  \, + \,  m_{_{\text{D}}} \right) \xi ^{\text{phys}}  \, = \, 0, \quad   \left[\xi^{\text{phys}} \right]\epsilon\gamma_{5}  \left(-\overleftarrow{\slashed{\partial} } \, + \,  m_{_{\text{D}}}  \right)   \, = \, 0 \ ,
    \label{DoubleDiracEquation}
\end{IEEEeqnarray}
where $m_{_{\text{D}}}$ stands for the (degenerate) supermultiplet mass.
The propagator for this field reads
\begin{IEEEeqnarray}{rl}
              \left(-\slashed{\partial} + m_{_{\text{D}}} \right)_{\alpha\alpha'} \left(-\slashed{\partial} + m_{_{\text{D}}} \right)_{\beta\beta'} \Delta_{F}\left( x\right) \ . 
     \label{Propagator}
 \end{IEEEeqnarray} 
 We use  the  Clebsh-Gordan coefficients to write  the states in the representation $ \frac{1}{2}\otimes\frac{1}{2}  $ as a sum  of  terms of definite zero and one   spins (and similarly for the antiparticle states). The  free tensor-spinor field can be  expressed in terms of the fields $ \phi $  and $ a_{\mu} $ as
 \begin{IEEEeqnarray}{rl}
           \xi ^{\text{phys}}  \left(  \sqrt{2}\epsilon\gamma_{5}\right) \, = \,   \left(  m_{_{\text{D}}} \, \phi -\gamma^{\mu} \partial_{\mu} \phi\right)  \gamma_{5}  \, + \,i\left( m_{_{\text{D}}} a_{\mu}\gamma^{\mu} - \tfrac{1}{4}f_{\mu\nu}\left[\gamma^{\mu},\gamma^{\nu} \right] \right) \ ,
    \label{ExpansionPropagatingFields}
\end{IEEEeqnarray}
with $ f^{\mu\nu}  \, = \,   \partial ^{\mu}a^{\nu}\, - \,  \partial ^{\nu}a^{\mu} $ (more details  on the derivation  of  this equation can be found in  \ref{Sec:ConventionsNotation}).  
 Extending  \eqref{ExpansionPropagatingFields} for the  off-shell case, the imposition of  \eqref{DoubleDiracEquation} gives:
\begin{itemize}
 \item[-] The Klein-Gordon equation for the   $ \phi $  field
\begin{IEEEeqnarray}{rl}
       \left(\square \, - \,m_{_{\text{D}}}^{2} \right)\phi \, = \, 0 \ .
    \label{Result}
\end{IEEEeqnarray} 
 \item[-] The Lorentz condition, the Proca equations and Bianchi identities for the  $ a_{\mu} $  field
\begin{IEEEeqnarray}{rl}      
        \partial^{\mu} a_{\mu }  \, = \, 0, \quad \partial^{\mu}f_{\mu\nu} \, - \,m_{_{\text{D}}}^{2}  a_{\nu }  \, = \, 0, \quad  \epsilon^{\mu\nu\rho\sigma}\partial_{\nu}f_{\rho\sigma}  \, = \,  0  \ .
    \label{LorentzEqMotionBianchi}
\end{IEEEeqnarray}
\end{itemize}

Having  identified  the  physical fields,  we write the general off-shell $ \xi $ field  (needed for supersymmetry) as the sum of its  symmetric and antisymmetric parts:
\begin{IEEEeqnarray}{rl}
            \xi \, = \, \xi_{s}  \, + \, \xi_{a} \ .
    \label{EsoEsTodo1}
\end{IEEEeqnarray}
In the covariant $ 4\times 4  $ basis $ \left(I,\gamma_{5},\gamma_{\mu},\gamma_{\mu}\gamma_{5},\left[\gamma_{\mu},\gamma_{\nu} \right]  \right)  $,  $ \xi_{s}  $  and  $ \xi_{a} $ are given by
\begin{IEEEeqnarray}{rl}
         \sqrt{2} \,\xi_{a} &\, = \,\left( i d\, - \, m_{_{\text{D}}}\phi\,\gamma_{5}  \, + \,      \left( b_{\mu}  \, + \,\partial_{\mu}\phi\right) \gamma^{\mu}\gamma_{5}\right)  \left[ \epsilon\gamma_{5}\right]   \ , \nonumber \\
          \sqrt{2}\, \xi_{s} &\, = \,i\left(- m_{_{\text{D}}} a_{\rho} \gamma^{\rho} \, + \, \tfrac{1}{4}\left( f_{\rho\sigma}  \, + \, c_{\rho\sigma}\right) \left[ \gamma^{\rho} ,\gamma^{\sigma} \right]\right)  \left[ \epsilon\gamma_{5}\right] \ .        
    \label{OffShellTensorCovBasis}
\end{IEEEeqnarray}
From \eqref{ComponentSUSYTransformations} we can read the transformation rules of the fields in \eqref{OffShellTensorCovBasis} under supersymmetry infinitesimal transformations
\begin{IEEEeqnarray}{rl}  
  \delta d    & \, = \,-i\tilde{\vartheta}\cdot \gamma_{5}\left[ \slashed{\partial}   \omega^{+}_{_{1}}   \, + \,   \omega^{+}_{_{2}}  \right], \nonumber \\   
      \delta         \phi     &  \, =\, -\tfrac{1}{m_{_{\text{D}}}} \tilde{\vartheta}\cdot \omega^{+}_{_{2}} ,      \nonumber \\ 
   \delta  a_{\mu}        &   \, = \, - \tfrac{i}{m_{_{\text{D}}} } \tilde{\vartheta}\cdot \gamma_{5} \left[ \partial_{\mu}  \omega^{-}_{_{1}}  \, + \,  \gamma_{\mu}  \omega^{-}_{_{2}}  \right],  \nonumber \\                 
       \delta        b_{\mu} &   \, = \, \tilde{\vartheta}\cdot\left\lbrace \tfrac{1}{2}\left[ \gamma_{\mu} ,\slashed{\partial}\right] \omega^{-}_{_{1}}  \, + \,\gamma_{\mu}\omega^{-}_{_{2}}  \, + \, \tfrac{1}{m_{_{\text{D}}}} \partial_{\mu} \omega^{+}_{_{2}}   \right\rbrace, \nonumber \\           
          \delta  c_{\mu\nu}   &   \, = \,i \tilde{\vartheta}\cdot \gamma_{5}\left( i \epsilon_{\mu\nu\rho\sigma}\partial^{\rho}\gamma^{\sigma}\gamma_{5}\omega^{+}_{_{1}}  \, - \,   \tfrac{1}{2}\left[ \gamma_{\mu},\gamma_{\nu} \right] \omega^{+}_{_{2}}  \, + \, \tfrac{1}{m_{_{\text{D}}}}\left[ \partial_{\mu},\gamma_{\nu} \right]\omega^{-}_{_{2}}\right),           
    \label{EsTocho}
\end{IEEEeqnarray}
with
\begin{IEEEeqnarray}{rl}
            \omega_{_{1}}^{\pm} \, = \,    \left( \chi^{\pm}\right)_{+}  \, +\, \left( \chi^{\mp}\right)_{-}, \quad       2\omega_{_{2}}^{\pm} \, = \,    \left( \mathcal{G}^{\pm}\right)_{+}  \, +\, \left( \mathcal{G}^{\mp}\right)_{-}\ , 
    \label{Omega}
\end{IEEEeqnarray}
and  $ \left( \cdots \right)_{\pm}   \, = \,  \frac{1}{2}\left(  I \pm \gamma_{5}\right) \left( \cdots \right) $. 
Next we ask what kinds of operators,  written in terms of $ \xi $, can generate the  free Lagrangian for the massive  scalar and vector fields, and algebraic equations  of motion for  the auxiliary fields $ d, b_{\mu} $ and  $ c_{\mu\nu}$.  The answer is
\begin{IEEEeqnarray}{rl}
   \mathcal{L}^{\xi}_{0}        \, = \,    \, - \,\tfrac{1}{4m_{_{\text{D}}}}\,\bar{\xi}\left(\slashed{\partial}\xi  \, + \, \xi\overleftarrow{\slashed{\partial}}^{\intercal}   \, + \,   2m_{_{\text{D}}} \xi \right) \ .
    \label{XiLagrangian}
\end{IEEEeqnarray}
To prove this, we first introduce the Lagrangians 
\begin{IEEEeqnarray}{rl}
             \mathcal{L}^{\xi_{s}}_{0}  &  \, \equiv \,  - \frac{1}{2m_{_{\text{D}}}} \bar{\xi}_{s}\left(m_{_{\text{D}}} +\slashed{\partial} \right)  {\xi}_{s} \ ,   \\
 \mathcal{L}^{\xi_{a}}_{0}  &  \, \equiv \,  - \frac{1}{2m_{_{\text{D}}}} \bar{\xi}_{a}\left(m_{_{\text{D}}} +\slashed{\partial} \right)  {\xi}_{a}  \ , 
    \label{QuePedo1}
\end{IEEEeqnarray}
that decompose in  the covariant $ 4\times 4 $  basis as (up to total derivative terms)
\begin{IEEEeqnarray}{rl}
      \mathcal{L}^{\xi_{a}}_{0}      & \, = \,  d^{*} d  \, + \ b^{*}_{\mu}    b^{\mu}   \, - \, m_{_{\text{D}}}^{2}\phi^{*}\phi   \, - \, \partial_{\mu}\phi^{*}\partial^{\mu}\phi  \ ,
    \label{QuePedo2}
\end{IEEEeqnarray}
\begin{IEEEeqnarray}{rl}
    \mathcal{L}^{\xi_{s}}_{0}        & \, = \,   \, - \,\tfrac{1}{2}f^{*}_{\mu\nu}f^{\mu\nu}   \, - \,m_{_{\text{D}}}^{2} a^{*}_{\mu}a^{\mu}   \, + \, \tfrac{1}{2}  c^{*}_{\mu\nu} c^{\mu\nu}  \ .
    \label{QuePedo3}
\end{IEEEeqnarray}
These are the required Lagrangians for the scalar and massive vector fields, with the  auxiliary fields completely decoupled from the propagating ones. Finally we note that their sum is equal to the expression  \eqref{XiLagrangian}.

\section{The free supersymmetric Lagrangian}
\label{sec:SupLagrangians}

Once we have described the bosonic sector of the superfields $ \Psi^{\pm} $, we  look for the form of the supersymmetric free  Lagrangian.  We introduce the following $ \mathcal{D}$-terms  (defined as the expansion coefficient of $  \left( \frac{1}{2}\right) \left( \vartheta\cdot\vartheta_{+}\right) \left( \vartheta\cdot\vartheta_{-}\right) $, up  to total derivatives):
\begin{IEEEeqnarray}{rl}
              \tfrac{1}{2} \left[ \bar{\Psi}^{-\varepsilon}  \Psi^{\varepsilon} \right]_{\mathcal{D}}    
                   &\, = \,  -\partial_{\mu}\bar{\chi}^{\varepsilon} \partial^{\mu}\chi^{\varepsilon}  \, - \,\partial_{\mu}\bar{\chi}^{(-\varepsilon)} \partial^{\mu}\chi^{-\varepsilon}  \, - \, \tfrac{1}{2}\bar{\xi}\slashed{\partial}\left(I  \, + \, \varepsilon\gamma_{5} \right) \xi  \nonumber \\
           &  \, + \,  \bar{\mathcal{G}}^{\varepsilon}\mathcal{G}^{\varepsilon}    \, - \, \bar{\chi}^{(-\varepsilon)}\overleftarrow{\slashed{\partial} } \mathcal{G}^{\varepsilon} \, + \, \bar{\mathcal{G}}^{\varepsilon}\slashed{\partial}\chi^{-\varepsilon}  \ ,
    \label{Dterm}
\end{IEEEeqnarray}
with $ \varepsilon  = \pm  $.  There are $ \mathcal{F}_{\pm} $-terms  that have the same dimensionality of \eqref{Dterm}:
\begin{IEEEeqnarray}{rl}
             \left[  \bar{\Psi}^{\varepsilon} \slashed{\partial}   \Psi^{\varepsilon}\right]_{\mathcal{F}_{\varepsilon}}        & \, = \, -\partial_{\mu}\bar{\chi}^{\varepsilon} \partial^{\mu}\chi^{\varepsilon}  \, - \,\partial_{\mu}\bar{\chi}^{-\varepsilon} \partial^{\mu}\chi^{(-\varepsilon)}  \, - \,  \bar{\chi}^{-\varepsilon}\overleftarrow{\slashed{\partial}}  \mathcal{G}^{\varepsilon} \nonumber \\
        &         \, + \,   \bar{\mathcal{G}}^{-\varepsilon} \slashed{\partial}  \chi^{\varepsilon} \, + \, \tfrac{1}{2}\bar{\xi}\left(I  \, + \,  \varepsilon\gamma_{5} \right)\left[ \xi\overleftarrow{{\slashed{\partial}} } ^{\intercal}  \right]   \ .
    \label{ComoAsi2}
\end{IEEEeqnarray}
The other bilinear at our disposal is 
\begin{IEEEeqnarray}{rl}
              \left[  \bar{\Psi}^{\varepsilon}    \Psi^{\varepsilon}\right]_{\mathcal{F}_{\varepsilon}}     & \, = \,  \bar{\chi}^{-\varepsilon} \mathcal{G}^{\varepsilon}  \, + \, \bar{\chi}^{-\varepsilon}\slashed{\partial}\chi^{-\varepsilon} \, + \,  \bar{\mathcal{G}}^{-\varepsilon} \chi^{\varepsilon}  \, +\, \bar{\chi}^{\varepsilon}{\slashed{\partial} }\chi^{\varepsilon}  \, + \, \tfrac{1}{2}\bar{\xi}\left(I  \, + \, \varepsilon\gamma_{5} \right) \xi    \ .    \label{AquiTodoEBum}
\end{IEEEeqnarray}
  
We construct the supersymmetric Lagrangian by taking  linear combinations of these three bilinears  and we adjust the constants by requiring, besides Hermiticity,  the absence of terms of the form $ -\partial_{\mu}\bar{\chi}^{\varepsilon} \partial^{\mu}\chi^{\varepsilon} $,  and the emergence of the free Lagrangians for the propagating fields,  when the auxiliary fields are  evaluated at their equations of motion. 
 The resulting free action is\footnote{In our conventions
\begin{IEEEeqnarray}{rl}
                \int d^{4}x[\cdots]_{D}  \, = \, -\frac{1}{2}\int d^{4}x d^{4}\vartheta [\cdots] , \quad \int d^{4}x[\cdots]_{\mathcal{F}_{\pm}}  = \pm \frac{1}{2}\int d^{4}x d^{4}\vartheta \delta^{2}(\vartheta_{\mp})[\cdots ] \nonumber 
     \label{FootNote}
 \end{IEEEeqnarray}
 with  $ \delta^{2}(\vartheta_{_{\pm}}) = \frac{1}{2}\vartheta^{\intercal}\epsilon\vartheta_{\pm}  $. }
\begin{IEEEeqnarray}{rl}
   A^{\text{Dirac}}_{0}  &  \, = \,     -\tfrac{1}{16m_{_{\text{D}}}}\int d^{4}x\,d^{4}\vartheta \left[ \bar{\Psi}^{-}  \Psi^{+}   \, + \, \bar{\Psi}^{+}  \Psi^{-} \right.  \nonumber \\
   & \left. \qquad \qquad \, + \, 2\delta^{2}\left(\vartheta_{-} \right)  \,  \bar{\Psi}^{+} \left( \slashed{\partial}   \, + \,  2m_{_{\text{D}}}  \right) \Psi^{+}   \, + \, \text{h.c.} \right]  
    \label{FreeSuperDiracAction}
\end{IEEEeqnarray}
with the corresponding Lagrangian density
\begin{IEEEeqnarray}{rl}
   \mathcal{L}^{\text{Dirac}}_{0}  &  \, = \,     \tfrac{1}{4m_{_{\text{D}}}}\sum^{+,-}_{\varepsilon}\left(\tfrac{1}{2} \left[ \bar{\Psi}^{-\varepsilon}  \Psi^{\varepsilon} \right]_{\mathcal{D}}  \, - \, \left[  \bar{\Psi}^{\varepsilon} \left( \slashed{\partial}   \, + \,  2m_{_{\text{D}}}  \right) \Psi^{\varepsilon}\right]_{\mathcal{F}_{\varepsilon}}  \right) \  \nonumber \\  
   & \, = \, \mathcal{L}^{\chi}_{0}  \, + \, \mathcal{L}^{\xi}_{0} \ ,
    \label{SuperSymmetricFreeLagrangian}
\end{IEEEeqnarray}
where $ \mathcal{L}^{\chi}_{0}$ stands for
\begin{IEEEeqnarray}{rl}
 \mathcal{L}^{\chi}_{0}   \, = \, \sum^{+,-}_{\varepsilon}\left( - \bar{\chi}^{\,\varepsilon}\,\slashed{\partial}\chi^{\varepsilon} \, + \,\tfrac{1}{4m_{_{\text{D}}}}\bar{\mathcal{G}}^{\varepsilon}\mathcal{G}^{\varepsilon} \, - \,\tfrac{1}{2}\bar{\mathcal{G}}^{\varepsilon}\,\mathcal{\chi}^{-\varepsilon}     \, - \,\tfrac{1}{2}\bar{\chi}^{(-\varepsilon)}\,\mathcal{G}^{\varepsilon} \right)  \ ,
    \label{ToOqueEuQuerias}
\end{IEEEeqnarray}
and  $ \mathcal{L}^{\xi}_{0}  $ is given by equation \eqref{XiLagrangian}, that is, the complete free Lagrangian reads:
\begin{IEEEeqnarray}{rl}
    \mathcal{L}^{\text{Dirac}}_{0}   & \, = \,    \, - \, m_{_{\text{D}}}^{2}\phi^{*}\phi   \, - \, \partial_{\mu}\phi^{*}\partial^{\mu}\phi   \,  \,  \, - \,\tfrac{1}{2}f^{*}_{\mu\nu}f^{\mu\nu}   \, - \,m_{_{\text{D}}}^{2} a^{*}_{\mu}a^{\mu}   \, - \,\bar{\chi}^{\,+}\,\slashed{\partial}\chi^{+}  \, - \,\bar{\chi}^{\,-}\,\slashed{\partial}\chi^{-}   \nonumber \\
      &\qquad  \, + \,\tfrac{1}{4m_{_{\text{D}}}}\left( \bar{\mathcal{G}}^{+}\mathcal{G}^{+}   \, + \,  \bar{\mathcal{G}}^{-}\mathcal{G}^{-} \right)  \, + \, d^{*} d  \, + \ b^{*}_{\mu}    b^{\mu} \, + \, \tfrac{1}{2}  c^{*}_{\mu\nu} c^{\mu\nu} \nonumber \\
     &\qquad    \, - \,\tfrac{1}{2}\left( \bar{\mathcal{G}}^{+}\,\mathcal{\chi}^{-}     \, + \,\bar{\chi}^{-}\,\mathcal{G}^{+}   \, + \,\bar{\mathcal{G}}^{-}\,\mathcal{\chi}^{+}     \, + \,\bar{\chi}^{+}\,\mathcal{G}^{-}   \right)\,. 
    \label{DiracCompleteComponentLagrangian}
\end{IEEEeqnarray}
The equations of motion for the  auxiliary fields  are
\begin{IEEEeqnarray}{rl}
            \mathcal{G}^{\pm}    \, = \, 2m_{_{\text{D}}} \chi^{\mp}, \quad d  \, = \, 0,\quad  b_{\mu}  \, = \, 0, \quad c_{\mu\nu}  \, = \, 0 \ .
    \label{VerdadePrincipal}
\end{IEEEeqnarray}
Inserting this solutions, in Eq. \eqref{DiracCompleteComponentLagrangian}, we obtain the Lagrangian for the scalar, Dirac and vector massive free fields.
Alternatively,  we can start  writing the supersymmetric equations of motion, that follow from the action \eqref{FreeSuperDiracAction}(with the definition $ \mathcal{D}_{\pm}^{2}  \, = \, \frac{1}{2}\mathcal{D}^{\intercal}\epsilon\mathcal{D}_{\pm} $):
\begin{IEEEeqnarray}{rl}
           \tfrac{1}{2} \mathcal{D}^{2}_{\pm} \Psi^{\pm}   \, \pm \, \left( \slashed{\partial}  \, + \,  2m_{_{D}} \right) \Psi^{\mp}  & \, = \, 0   \, .
    \label{FreeSUSYEqMotion}
\end{IEEEeqnarray}
With the help of $ \frac{1}{4}\mathcal{D}^{2}_{\mp}\mathcal{D}^{2}_{\pm} \Psi^{\pm}  \, = \, -\square \Psi^{\pm}$, we can see that these motion equations hold if and only if  the superfields $ \Psi^{\pm} $ are simultaneous solutions of the Dirac and the ``Wess-Zumino'' equations:
\begin{IEEEeqnarray}{rl}
              \left( \slashed{\partial}  \, + \,  m_{_{D}} \right) \Psi^{\pm}  \, = \,0 , \quad \tfrac{1}{2}\mathcal{D}^{2}_{\pm} \Psi^{\pm}  \, = \, \mp m_{_{\text{D}}}\Psi^{\pm}   \ .
     \label{DiracSuperfieldEquationsOfMotion}
 \end{IEEEeqnarray} 
Thus, the Dirac superfield in the interaction picture reads
 \begin{IEEEeqnarray}{rl}
            \left(  {\Psi}^{\pm} \right)^{\text{free}} & \, =\, \left( \chi^{\pm} \right)^{\text{free}}   \, - \,  \sqrt{2} \vartheta_{\pm}\cdot \gamma_{5} \xi^{\text{free}} \,+ \, m_{_{\text{D}}} \left( \vartheta\cdot \vartheta_{\pm}\right) \left( \chi^{\mp} \right)^{\text{free}} \ ,  
     \label{DefFreeDiracSuperField}
 \end{IEEEeqnarray}
 where  $ \left( \chi^{\pm} \right)^{\text{free}} $ satisfies the Dirac equation and $ \xi^{\text{free}}  $ is  of the form \eqref{ExpansionPropagatingFields}, satisfying \eqref{DoubleDiracEquation}.  These interaction-picture superfields have been obtained in \cite{Jimenez:2014gfa} by the noncanonical methods of Weinberg.

It is instructive to examine the spinor superfields and the Lagrangian contained in  \cite{Siegel:1979ai} (Siegel model) within  the context of the present work. In  Siegel model, only the right and left superfields $ \left( \Phi^{+}\right)_{+\alpha}  $ and  $ \left( \Phi^{-}\right)_{-\alpha}  $  are introduced (including their adjoints), 
\begin{IEEEeqnarray}{rl}
             \left(  {\Phi}^{\pm} \right)_{\pm \alpha} & \, =\, \left( \chi^{\pm} \right) _{\pm\alpha}   \, + \,   \left( \gamma_{5}\vartheta\right) _{\pm\alpha}\left( \xi_{0}  \, \pm  \, \xi_{5}\right)  \nonumber \\
           &\qquad  \, - \,\tfrac{1}{4}\xi_{\mu\nu}\left( \left[\gamma^{\mu} , \gamma^{\nu}\right] \gamma_{5}\vartheta\right) _{\pm\alpha}\, + \, \vartheta\cdot  \vartheta_{\pm}\,\left(  \mathcal{G}^{\pm}  \, + \, \slashed{\partial}\chi^{\mp} \right)_{\pm\alpha} \ .  
    \label{Siegel-01}
\end{IEEEeqnarray}
In writing the  tensor $ \xi_{\pm\alpha,\pm\beta}   $, we have used the  fact that  only  the projection $ (0,0)^{2}\oplus\left[ (1,0)\otimes (0,1)\right] $  of the Lorentz group appears. The fields $ \xi_{0} $,  $ \xi_{5} $ and  $ \xi_{\mu\nu} $  correspond  to the coefficients of $ I,\gamma_{5} $ and  $ \frac{1}{4}\left[\gamma^{\mu} \gamma^{\nu}\right] $ in the $ 4\times 4 $ covariant basis, respectively.  For the massive case, just by counting degrees of freedom, we can see that all component fields are propagating except for one scalar field.
The  free supersymmetric action  of Siegel model is written as follows
\begin{IEEEeqnarray}{rl}
       {A}_{\text{Siegel}}  \, = \,   \tfrac{1}{4}\int d^{4}x d^{4}\vartheta \,  G^{*}G  \, - \, \left( { m^{2}_{_{\text{D}}}}/{2}\right) \int d^{4}x \,d^{2}\vartheta_{+}\,\bar{\Phi}^{+}  {\Phi}^{+}_{\,\,+}   \, + \, \text{h.c.}
    \label{Siegel-02}
\end{IEEEeqnarray}
where 
\begin{IEEEeqnarray}{rl}
             G   &\, = \, \mathcal{D}\cdot\left(\Phi^{+}  \, + \, \Phi^{-} \right)\, = \, \left(\mathcal{D}_{+}\cdot\Phi^{+}_{\,\,+}  \, + \, \mathcal{D}_{-}\cdot\Phi^{-}_{\,\,-} \right)\ .
    \label{Siegel-03}
\end{IEEEeqnarray}
In terms of the  fermion fields 
\begin{IEEEeqnarray}{rl}
            \eta^{+}_{\alpha}  \, = \, \sqrt{2} m_{_{\text{D}}}\left( \chi^{+}\right) _{+\alpha}  \, + \, \tfrac{1}{\sqrt{2}}\left( \mathcal{G}^{+}\right) _{-\alpha}   , \quad    \eta^{-}_{\alpha}  \, = \, \sqrt{2} m_{_{\text{D}}}\left( \chi^{-}\right) _{-\alpha}  \, + \, \tfrac{1}{\sqrt{2}}\left( \mathcal{G}^{-}\right) _{+\alpha}\   , \nonumber\\
    \label{Siegel-04}
\end{IEEEeqnarray}
and  the bosonic fields
\begin{IEEEeqnarray}{rl}
       \xi_{0}  \, = \, \tfrac{1}{2} D, \quad    \xi_{5}  \, = \, \tfrac{1}{2} \,A,\quad \xi_{\mu\nu}  \, = \, \tfrac{1}{2\sqrt{2}}\,\tilde{B}_{\mu\nu} \ ,
    \label{Siegel-05}
\end{IEEEeqnarray}
the supersymmetric action \eqref{Siegel-02}, expressed in components, reads:
\begin{IEEEeqnarray}{ll}
       {A}_{\text{Siegel}}  \, = \,   (-) \int d^{4}x\left[  \bar{\eta}^{+}\left( \slashed{\partial}   \, + \,  m_{_{\text{D}}} \right) \eta^{+}  \, + \,\bar{\eta}^{-}\left( \slashed{\partial}   \, + \,  m_{_{\text{D}}} \right)  \eta^{-} \right]\nonumber \\
          \qquad  \, + \,    \int d^{4}x\left[ \partial^{\mu}\tilde{B}^{*}_{\mu\nu}\,\partial_{\rho}\tilde{B}^{\rho\nu}   \, + \,  m^{2}_{_{\text{D}}} \tilde{B}^{*}_{\mu\nu}\tilde{B}^{\mu\nu}  \, -\,\partial_{\mu}A^{*}\partial^{\mu}A \, - \,  m^{2}_{_{\text{D}}} A^{*}A   \, + \,D^{*}D\right] \ .\nonumber \\
    \label{Siegel-06}
\end{IEEEeqnarray} 
 Therefore, one of the main differences between Lagrangians  \eqref{FreeSuperDiracAction} and \eqref{Siegel-06}, is that in the former case, the spin one particle is propagating through a vector field,  while in the later case, through a second-rank  antisymmetric tensor. We can see that the free  superfields $\left(   \Phi^{\pm}_{\pm}\right) ^{\text{free}} $  are in agreement with the free superfields reported in \cite{Jimenez:2014gfa}. 
 

Written in terms of the dual tensor field $ {B}_{\mu\nu}   \, = \,  \frac{1}{2}\epsilon^{\mu\nu\rho\sigma}\tilde{B}_{\mu\nu} $, the Lagrangian  for the spin-one particle is~\cite{Kalb:1974yc}:
\begin{IEEEeqnarray}{rl}  
 \int d^{4}x \left( \partial^{\mu}\tilde{B}^{*}_{\mu\nu}\,\partial_{\rho}\tilde{B}^{\rho\nu}   \, + \,  m^{2}_{_{\text{D}}} \tilde{B}^{*}_{\mu\nu}\tilde{B}^{\mu\nu} \right)  &\, = \,  \nonumber \\    
                  - \int d^{4}x  &\left(  \frac{1}{6}F_{\mu\nu\rho}^ {*}   F^{\mu\nu\rho}  \, + \,  {m^{2}_{_{\text{D}}}}{B}^{*}_{\mu\nu}{B}^{\mu\nu} \right)  \ ,
    \label{2AntiSymTensor-MassiveAction}
\end{IEEEeqnarray} 
with
\begin{IEEEeqnarray}{rl}
            F^{\mu\nu\rho}  \, = \,  \partial^{\mu}B^{\nu\rho}  \, + \, \partial^{\nu}B^{\rho\mu} \, + \, \partial^{\rho}B^{\mu\nu}\ .
    \label{RamondKalb_3Form}
\end{IEEEeqnarray}

 The introduction of the right and left superfields $ \left( \Psi^{+}\right)_{-\alpha}  $ and  $ \left( \Psi^{-}\right)_{+\alpha}  $   in the Lagrangian \eqref{FreeSuperDiracAction}, enables us to express the particles of the supermultiplet as the lowest mass dimension fields (scalar, Dirac and vector), at the expense of introducing more auxiliary fields (including fermionic).  
 
 The model in  \cite{Siegel:1979ai}  describes a 8 + 8 realization of supersymmetry, whereas our model (due to the extra superfields) constitutes a 16 + 16 realization. An analogous analysis can be made  for the model in \cite{Gates:1980az} (which has the same number of off-shell fermionic degrees of freedom as \cite{Siegel:1979ai}), in this case, the spin particle is propagating through a vector field and the spin zero particle through 3-form field $ (\Gamma^{\mu\nu\rho}) $. 
 
Before closing this section, we comment on the other approaches that do not make use of chiral superfields. There are very well known models for superspin one-half in which the component propagating fields have  minimal mass dimensions~\cite{Siegel:1979ai,VanProeyen:1979ks,Mukhi:1979wc}. They are based on unconstrained scalar superfields (usually taken to be real).  Recent works constructed with these type of superfields include the introduction of auxiliary superfields, the superspin one and three-halves cases, and models with complex mass parameters \cite{Buchbinder:2002gh,Buchbinder:2002tt,Gregoire:2004ic,Gates:2013tka, D'Auria:2004sy,Louis:2004xi,Theis:2004pa,Kuzenko:2004tn} (For an alternative gauge formulation of the massive superspin-1/2 multiplet, see \cite{Gates:2005su}). Among their component fields, these models feature the vector field for the spin one particle, and consequently they are known as the  vector multiplet models (tensor multiplet for case of the three-half superspin). The simplest prototype of these kind of models is based  on a unconstrained superfield $ V $~\cite{Siegel:1979ai,VanProeyen:1979ks,Mukhi:1979wc}. In this case, three of the four scalar component fields  of $ V $ are auxiliary, and the rest of the components are propagating fields. The models containing the scalar superfield $ V $ and those built with the chiral spinor superfields $ \Phi^{\pm}_{\pm} $ are  equivalent  even in the presence  of supergravity \cite{Siegel:1979ai,VanProeyen:1979ks,Mukhi:1979wc}. However, as far as we know, it is not clear if this duality holds for arbitrary  potentials of $ V $. 
\section{Superpropagators}
\label{sec:Superpropagators}
On general grounds, we would expect  the couplings of the  antisymmetric propagating field $ B^{\mu\nu} $ of Siegel massive model to possess one degree more of UV divergence  with respect to the propagating vector field $ a^{\mu} $ of the Dirac superfields, but in a supersymmetric theory, we have to keep in mind that the  auxiliary superfields also contribute to the correlation functions in superspace.  In this section,  we show that  not only the superpropagators  of the two mentioned models (with actions \eqref{FreeSuperDiracAction} and \eqref{Siegel-02}) possess the same degree of divergence, but (in essence) they coincide. 

An attractive feature of  models with superfields that are only the restricted by their supersymmetric chiral condition is the fact that path integrals can be easily implemented. Upon the  identification
\begin{IEEEeqnarray}{rl}
            \left(  \Psi^{\pm}\right)_{\alpha}   \, = \, \mathcal{D}^{2}_{\mp} \left( \mathcal{S}^{\pm}\right)_{\alpha}, \quad \bar{\Psi}^{\pm}  \, = \, -\mathcal{D}^{2}_{\mp} \left( \bar{\mathcal{S}}^{\pm}\right)_{\alpha} \ ,
    \label{PsiAsDerOFGenSuperFields}
\end{IEEEeqnarray}   
 we can integrate over a set of four-spinor general superfields $ \Pi_{\alpha}  \, = \, (\mathcal{S}^{+},\mathcal{S}^{-}) _{\alpha}$, without further constraining  the integral functional \cite{Weinberg:2000cr}.   The corresponding Green functions can be extracted from  functional integrals of the form
\begin{IEEEeqnarray}{rl}
           \frac{1}{\text{Const.}} \int \left[  \prod_{c} d\bar{\Pi}_{c} d\Pi_{c}\right] \left( \bar{\Pi}_{a_{_{1}}}\dots\bar{\Pi}_{a_{_{n}}}{\Pi}_{b_{_{1}}}\dots{\Pi}_{b_{_{n}}}\right) \exp{\left[ -i \sum_{ab}D_{ab}\Pi^{*}_{a}\Pi_{b}\right] } \ ,
    \label{CadaSemestre}
\end{IEEEeqnarray}
with $ a,b,a_{_{1}},\dots , b_{_{n}} $ running over continuous and discrete indices.  The free action is always invariant under 
\begin{IEEEeqnarray}{rl}
              \delta \Pi_{\alpha}  \, = \,\sum_{\beta} \left( \mathcal{D}_{+\beta}\,\eta_{\beta\alpha} ,\mathcal{D}_{-\beta}\,\tau_{\beta\alpha}\right) \  ,
     \label{DelPiChang}
 \end{IEEEeqnarray} 
 for arbitrary superspace functions  $ \eta_{\beta\alpha} $ and $ \tau_{\beta\alpha} $. Thus,  the superpropagator $\Delta_{ab}\equiv \langle \Pi_{a} \bar{\Pi}_{b}\rangle$ satisfies the relation 
 $ \sum_{c}D_{ac}\Delta_{cb}   \, = \, \mathcal{P}_{a}\delta_{ab}$ , where $ \delta_{ab} $ is a product of Dirac and Kronecker delta functions and $ \mathcal{P} $  is the chiral projection matrix
\begin{IEEEeqnarray}{rl}
            \mathcal{P}  \, = \, \frac{1}{-4 \square } \begin{pmatrix}
\mathcal{D}^{2}_{+}\mathcal{D}^{2}_{-} & 0 \\ 
0 & \mathcal{D}^{2}_{-}\mathcal{D}^{2}_{+} 
\end{pmatrix} 
    \label{DiracProjectionMatrix}\ .
\end{IEEEeqnarray}
Once the general structure of $ \Delta_{ab} $ has been determined, from Eq. \eqref{PsiAsDerOFGenSuperFields} we can straightforwardly obtain the superpropagators for the fields $  {\Psi}^{\pm} $ and $  \bar{\Psi}^{\pm} $.
In terms of the general superfields \eqref{PsiAsDerOFGenSuperFields},  the  free action \eqref{FreeSuperDiracAction}  reads
\begin{IEEEeqnarray}{rl}
        {A}^{\text{Dirac}}_{0}  & \, = \,    \tfrac{1}{16 m_{_{\text{D}}}} \int d^{4}x d^{4}\vartheta \left[\bar{\mathcal{S}}^{-} \mathcal{D}^{2}_{+} \mathcal{D}^{2}_{-}  \mathcal{S}^{+} \, + \, \bar{\mathcal{S}}^{+} \mathcal{D}^{2}_{-} \mathcal{D}^{2}_{+}  \mathcal{S}^{-} \right.   \nonumber \\
          &  \left.  \qquad       \, -\,  2\left(  \bar{\mathcal{S}}^{+} \left( \slashed{\partial}   \, + \,  2m_{_{\text{D}}}  \right) \mathcal{D}^{2}_{-} \mathcal{S}^{+}\right)     \, + \,  2\left(  \bar{\mathcal{S}}^{-} \left( \slashed{\partial}   \, + \,  2m_{_{\text{D}}}  \right) \mathcal{D}^{2}_{+} \mathcal{S}^{-}\right) \right].
    \label{ScalarKineticLagrangian1}
\end{IEEEeqnarray}
and the relevant superpropagators of the model become 
\begin{IEEEeqnarray}{rl}
    \left\langle\Psi_{\alpha}^{\pm}\left(z_{_{1}} \right)\, \bar{\Psi} _{\beta}^{\mp}\left(z_{_{2}} \right)  \right\rangle & \, = \,(-i)\left(-{\slashed{\partial}  \, + \, m_{_{\text{D}}}}{}\right)_{\alpha\beta}\Delta_{F}\left(x^{\pm}_{_{12}}\right)\ , \quad 
    \label{DiracSupPropagator1}
\end{IEEEeqnarray}
\begin{IEEEeqnarray}{rl}
     \left\langle\Psi_{\alpha}^{\pm}\left(z_{_{1}} \right)\, \bar{\Psi}_{\beta}^{\pm}\left(z_{_{2}} \right)   \right\rangle  & \, = \, \pm  (-i)\,2\delta^{2}\left[ \left( \vartheta_{_{1}}  \, - \, \vartheta_{_{2}} \right)_{\pm} \right]\nonumber \\
      &\quad \times \left\lbrace m_{_{\text{D}}}\left(-{\slashed{\partial} \, + \, m_{_{\text{D}}}}\right)_{\alpha\beta}\Delta_{F}\left(x^{\pm}_{_{12}}\right)      \, + \, \delta_{\alpha\beta}\delta\left(x^{\pm}_{_{12}}\right) \right\rbrace   \ ,\nonumber \\
    \label{DiracSupPropagator2}
\end{IEEEeqnarray}
with $ \Delta_{F}(x) $ as the Feynman massive propagator  and  
\begin{IEEEeqnarray}{rl}
            \left( x^{\pm}_{_{12}} \right)^{\mu}  \, = \, x^{\mu}_{_{1}} - x^{\mu}_{_{2}}  \, + \, \left(\vartheta_{_{2}} -\vartheta_{_{1}} \right)\cdot\gamma^{\mu}\left( \vartheta_{_{_{2}}\mp}+ \vartheta_{_{_{1}}\pm} \right) \ .
    \label{PairingPlusMinusPlusMinus}
\end{IEEEeqnarray}

 Notice that the superpropagators in \eqref{DiracSupPropagator1} and \eqref{DiracSupPropagator2} can be directly compared  with those of the  Wess-Zumino model, as both cases can be written in the form
\begin{IEEEeqnarray}{rl}
              (-i)P(-i\partial)\Delta_{F}\left(x^{\pm}_{_{12}}\right), \quad (\mp 2 i)\,\delta^{2}\left[ \left( \vartheta_{_{1}}  \, - \, \vartheta_{_{2}} \right)_{\pm} \right] \tilde{P}(-i\partial)\Delta_{F}\left(x^{\pm}_{_{12}}\right)\ , 
    \label{BothModelsWessDirac}
\end{IEEEeqnarray}
where the Wess-Zumino case is recovered through the relation $ P(-i\partial) =\tilde{P}(-i\partial)/m_{_{\text{S}}} =1$ while the Dirac case follows from
\begin{IEEEeqnarray}{rl}
              P(-i\partial)  \, = \, (-\slashed{\partial} + m_{_{\text{D}}}), \quad    \tilde{P}(-i\partial)   \, = \,  m_{_{\text{D}}}P(-i\partial)   \, + \,\left( m^{2}_{_{\text{D}}}  \, - \,\square\right)  \ . 
     \label{DiracPolynomials}
 \end{IEEEeqnarray} 
 
  Working in the same lines to obtain the superpropagators of the Siegel model (discussed at the end of Section \ref{sec:SupLagrangians}),  we find 
 \begin{IEEEeqnarray}{rl}
        \langle \Phi^{\pm}_{\pm\alpha}\left(z_{_{1}} \right)\bar{\Phi}_{\pm\beta}^{\pm}\left(z_{_{2}} \right) \rangle  & \, = \, \pm 2 \delta^{2}\left[ \left( \vartheta_{_{1}}  \, - \, \vartheta_{_{2}} \right)_{\pm} \right]\left(  2m^{2}_{_{\text{D}}}  \, - \, \square    \right)\delta_{\pm\alpha,\pm\beta}  \Delta_{F}\left(x^{\pm}_{_{12}}\right)  \nonumber \\
    \langle\Phi_{\pm\alpha}^{\pm}\left(z_{_{1}} \right)\bar{\Phi}_{\mp\beta}^{\mp}\left(z_{_{2}} \right) \rangle  & \, = \,\left( -\slashed{\partial} \right)_{\pm\alpha,\mp\beta} \Delta_{F}\left(x^{\pm}_{_{12}}\right)\ . 
    \label{Siegel_Superpropagators}
\end{IEEEeqnarray}
Recall that the subscripts  $ + $ and $ - $ represent left and right projections, respectively, and note that the left and right projections of the superpropagators \eqref{DiracSupPropagator1} and \eqref{DiracSupPropagator2} are the same as the superpropagators   \eqref{Siegel_Superpropagators}.

 We conclude this section pointing out that the  superpropagators  obtained in the present work are consistent with those reported in \citep{Jimenez:2014gfa}.

\section{Super Yukawa interactions}
\label{sec:SupYuk}
The purpose of this section is to outline the form of possible interaction potentials.  We introduce the scalar superfields $ \Omega_{+}  $  and $ \Omega_{-}  $ 
\begin{IEEEeqnarray}{rl}
        \Omega^{\pm} &  \, = \,  z^{\pm}   \, - \,  \sqrt{2} \vartheta_{\pm}^{\intercal}\epsilon\, \lambda\, + \, \vartheta\cdot  \vartheta_{\pm}\,  \mathcal{R}^{\pm}   \ ,
     \label{DefScalarSuperField}
 \end{IEEEeqnarray}
 with their corresponding free Lagrangian 
\begin{IEEEeqnarray}{rl}
            \mathcal{L}^{\text{Scalar}}_{0}  & \, = \, \tfrac{1}{2}\sum^{+,-}_{\varepsilon}\left[ \Omega^{*\varepsilon}  \Omega^{(-\varepsilon )} \right]_{\mathcal{D}}  \, - \,  m_{\text{s}} \sum^{+,-}_{\varepsilon} \left[  \Omega^{*(\varepsilon) }  \Omega^{\varepsilon}\right]_{\mathcal{F}_{\varepsilon}}   \nonumber \\
            &  \, = \, -\sum^{+,-}_{\varepsilon}\partial_{\mu}z^{\varepsilon *} \partial^{\mu}z^{\varepsilon}
      \, - \, \bar{\lambda}\slashed{\partial}\lambda  \, + \,\sum^{+,-}_{\varepsilon}\mathcal{R}^{\varepsilon *}\mathcal{R}^{\varepsilon} \nonumber \\
       &  \, -\,  m_{\text{s}}\left(\sum^{+,-}_{\varepsilon}\left( z^{\varepsilon *} \mathcal{R}^{-\varepsilon}\, + \, \mathcal{R}^{\varepsilon *}   z^{-\varepsilon} \right)   \, +\,  \bar{\lambda}  \lambda \right)  \ ,   
    \label{ScalarKineticLagrangian2}
\end{IEEEeqnarray}
and  $ \Omega^{*\mp} \left( x,\vartheta\right)  =\left[ \Omega^{\pm} \left( x,\vartheta\right) \right]^{*}   $. In order to construct the  trilinear superpotential for  scalar and  Dirac superfields, we have at our  disposal several combinations formed out of  $ \Omega_{+} $  and  $ \Omega^{*}_{+} $  and left-right projections  of the bilinears $ \left( \bar{\Psi}^{+}\right)_{\alpha} \left( {\Psi}^{+}\right)_{\beta}   $ and $ \left( {\Psi}^{+}\right)_{\alpha} \left( {\Psi}^{+}\right)_{\beta}   $. 
As an example we take 
\begin{IEEEeqnarray}{rl}
            \mathcal{W}& \, = \,h\,\Omega^{+}\bar{\Psi}^{+}\left[ \tfrac{1}{2} \left( I \, - \, \gamma_{5}\right) \right] \Psi^{+}  \ ,             
    \label{ExampleInteractionPotential}
\end{IEEEeqnarray}
whose $ \mathcal{F}_{+} $-term reads
\begin{IEEEeqnarray}{rl}
    \left[  \mathcal{W}\right]_{\mathcal{F}_{+}}  & \, = \,   
     h\,\mathcal{R}^{+}\overline{\chi^{-}}\left(  \chi^{+}  \right)_{-}  \, + \,      h\, {z}^{+}\left( \overline{\chi^{-}}\left( \mathcal{G}^{+} \, + \, \slashed{\partial}\chi^{-} \right)_{-} \, + \, \left( \bar{\mathcal{G}}^{-}  \, - \, \bar{\chi}^{+}\overleftarrow{\slashed{\partial}} \right)\left(  \chi^{+}\right) _{-}\right) \nonumber \\
     &\quad   \, + \, \tfrac{1}{4}\,  h\,  z^{+} \,\bar{\xi}\left(I+\gamma_{5} \right)  \xi \left(I-\gamma_{5} \right)    \, + \, h\, \bar{\xi}\,\left( \lambda\right) _{+}\left( \mathcal{\chi}^{+} \right)_{-}\, -\, \left( \epsilon\lambda\right)_{+} \left( \overline{\chi^{-}}\right)_{-} \xi  \ . \nonumber \\
        \label{ComoChingan}
\end{IEEEeqnarray}

In terms of the expansion for  $ \xi_{\alpha\beta} $ we have
\begin{IEEEeqnarray}{rl}
 \tfrac{1}{4} h\,\bar{\xi}\left(I+\gamma_{5} \right)  \xi \left(I-\gamma_{5} \right)     & \, = \,  -h\left(  i\,m_{_{\text{D}}} a^{*}_{\mu}  \, +\,   b^{*}_{\mu}  \, + \,\partial_{\mu}\phi^{*} \right) \left(    i\,m_{_{\text{D}}} a^{\mu}  \, +\,  b^{\mu}  \, + \,\partial^{\mu}\phi \right) \ , \nonumber \\   
        \sqrt{2}  \,\left( \lambda\right) _{+} \bar{\xi}\left( {\chi}^{+}\right)_{-}   &\, = \, \lambda _{+}\cdot \gamma_{5}\gamma^{\mu} \left( i\,m_{_{\text{D}}} a^{*}_{\mu} \gamma_{5}  \, + \,        b^{*}_{\mu}  \, + \,\partial_{\mu}\phi^{*}\right) \left( \chi^{+} \right)_{-}  \ , \nonumber \\
       \sqrt{2}  \left( \epsilon\lambda\right)_{+}\xi\, \overline{\left( {\chi}^{-}\right)}_{-}  &\, = \, \overline{\left( {\chi}^{-}\right)}_{-}\left(   - i\,m_{_{\text{D}}} a_{\mu}  \gamma_{5} \, + \,     b_{\mu}  \, + \,\partial_{\mu}\phi \right)\gamma^{\mu}  \left( \lambda\right)_{+} \ .
    \label{Siguieron}
\end{IEEEeqnarray}

The non-appearance of the strength tensor $ f_{\mu\nu} $ is only due  to the fact that we have chosen the left projection for the superfield $ \Psi^{+} $.  We note  that the algebraic solutions to the auxiliary fields become non-polynomial in the interacting theory. Although the introduced super Yukawa potential possesses non-renormalizable interactions,  we should expect that due to  non-renormalization theorems, their ultraviolet behavior will improve.
 The coupling \eqref{ExampleInteractionPotential} shows an explicit difference between  the model of Siegel and ours, this  interaction superpotential is not available  in the former case,  since only  left and right superfields $ \Phi^{-}_{-} $  and $ \Phi^{+}_{+} $ are introduced.

\section{Reality conditions}
\label{sec:RealityCon}
In considering  two superfields with opposite chirality for the scalar and Dirac cases,  we have  allowed for the whole supermultiplet to carry internal non trivial quantum numbers. To recover the usual Majorana field from the linear term  of the scalar superfield, we must impose the reality condition $ \Omega^{*\pm}   \, = \, \Omega^{\pm}  $, from which we have
\begin{IEEEeqnarray}{rl}
               \left( z^{+}\right) ^{*}  \, = \, z^{-} , \quad   \left( \mathcal{R}^{+}\right) ^{*}  \, = \,  \mathcal{R}^{-}, \quad \lambda  \, = \, -\epsilon\gamma_{5}\beta\lambda^{*}\ .
    \label{RealityConditionsScalar}
\end{IEEEeqnarray}
For the Dirac superfield, we impose the Majorana Condition
\begin{IEEEeqnarray}{rl}
           \bar{\Psi}^{\pm}_{\alpha}  \, = \,\left( \epsilon\gamma_{5} {\Psi}^{\pm}\right)_{\alpha}     
    \label{SuperMajoranaCondition}
\end{IEEEeqnarray}
that in terms of its components reads 
\begin{IEEEeqnarray}{rl}
            \left( {\chi}^{+} \right) ^{*} \, = \, -\epsilon\gamma_{5}\beta {\chi}^{-}, \quad   \left( \mathcal{G}^{+} \right) ^{*} \, = \, -\epsilon\gamma_{5}\beta \mathcal{G}^{-}, \quad  \xi \, = \, - \epsilon\gamma_{5}\,\bar{\xi}\, \epsilon\gamma_{5} \ .
    \label{MajoranaFermionSector}
\end{IEEEeqnarray}
This last condition implies the following relations for the bosonic fields
\begin{IEEEeqnarray}{rl}
          d  \, = \, d^{*}, \quad  \phi \, = \,\phi^{*}, \quad b_{\mu}\, = \, b_{\mu}^{*} , \quad
          a_{\mu}  \, = \, a_{\mu}^{*}, \quad   c_{\rho\sigma}  \, = \,  c^{*}_{\rho\sigma}\  .
    \label{MajoranaBosonSector}
\end{IEEEeqnarray}
Therefore, the bosonic propagating and auxiliary  fields become all real. Since any Dirac superfield can be written as the sum of two Majorana superfields,  expression \eqref{SuperSymmetricFreeLagrangian} splits into  two copies of the Lagrangian for  Majorana superfields.  
 
By identifying 
\begin{IEEEeqnarray}{rl}
             \Psi =\Psi^{+},\quad \bar{\Psi}=\bar{\Psi}^{-} ,\quad \Psi^{-}  \, = \, \bar{\Psi}\epsilon\gamma_{5}, \quad \bar{\Psi}^{+}  \, = \,-{\Psi}^{\intercal} \epsilon\gamma_{5}\ ,\nonumber\\
    \label{OneDiracMajorana}
\end{IEEEeqnarray}
 the action in Eq. \eqref{FreeSuperDiracAction} for the case of only one superspin one-half multiplet,  can be written entirely in terms of only the Dirac chiral superfield $ \Psi $ $ (\mathcal{D}_{+}\Psi =0) $ and its Dirac adjoint $ \bar{\Psi} $ ($ \mathcal{D}_{-}\bar{\Psi} =0 $) as follows:
\begin{IEEEeqnarray}{rl}
   A_{0}  &  \, = \,     -\tfrac{1}{16m_{_{\text{D}}}}\int d^{4}x\,d^{4}\vartheta \left[ \bar{\Psi}  \Psi \, - \, \delta^{2}\left(\vartheta_{-} \right)  \,  {\Psi}\cdot \left( \slashed{\partial}   \, + \,  2m_{_{\text{D}}}  \right) \Psi   \, + \, \text{h.c.} \right] \ . \nonumber\\
    \label{FreeSuperDiracActionOneSuperparticle}
\end{IEEEeqnarray}
 The corresponding component Lagrangian, is   \eqref{DiracCompleteComponentLagrangian} with the complex fields evaluated at \eqref{MajoranaFermionSector} and \eqref{MajoranaBosonSector}.

\section{Two Component Notation.}
\label{sec:twoComponent}
Through the manuscript, we have adopted the 4-spinor component notation of~\cite{Weinberg:2000cr}. In this section, we rewrite some of our findings in the corresponding 2-spinor dotted-undotted (Van der Waerden) notation, for the case of the Majorana chiral superfield. For every index $ \alpha =1,2,3,4 $, we introduce an undotted index $ (a=1,2) $  and dotted index  $ (\dot{a}=\dot{1},\dot{2}) $ to represent the right projection  $ (+\alpha) $ and  the left projection  $ (-\alpha) $ of the 4-spinor $ \alpha $, respectively. The 
4-spinor  variable $ \vartheta_{\alpha}   $ and the 4-spinor superderivative, in terms of dotted-undotted indices, are expressed as 
\begin{IEEEeqnarray}{rl}
            \vartheta_{\alpha}    \, = \,\begin{pmatrix}
\theta_{a} \\ 
\bar{\theta}^{\dot{a}}
\end{pmatrix}\ , \quad
             \mathcal{D}_{\alpha}  \, = \,\begin{pmatrix}
D_{a} \\ 
\bar{D}^{\dot{a}}
\end{pmatrix}.
     \label{7-01}
 \end{IEEEeqnarray} 
 The inner product between two undotted two-spinors $ v_{a} $  and  $ v'_{a} $  is given by
\begin{IEEEeqnarray}{rl}
             v v'  \, = \, v^{a}v'_{a}  \, = \, -v_{a}v'^{a} \ .
    \label{7-02}
\end{IEEEeqnarray}
We have used   $ e^{ab}$,  the two-dimensional totally antisymmetric tensor ($ e^{12}=+1 $), to raise the index of $ v'_{a} $. The tensor $ e_{ab} $, satisfies $  e_{ab}e^{bc}  \, = \, \delta_{a}^{c} $. Similar remarks apply for dotted-spinors.
 To convert  the four by four matrices $ \gamma^{\mu} $ and $ \Sigma^{\mu\nu}  \, = \, (-i/4)\left[ \gamma^{\mu},\gamma^{\nu}\right] $ into  two-dimensional notation,  we apply the following substitution rules:
\begin{IEEEeqnarray}{rl}
\gamma^{\mu}_{+\alpha-\beta}  & \,\, \rightarrow\,\, -i\left( \sigma^{\mu}\right) _{a\dot{b}} , \quad  \gamma^{\mu}_{-\alpha+\beta}   \, \,\rightarrow \,\, -i\left( \bar{\sigma}^{\mu}\right) ^{\dot{a}{b}} \nonumber \\
            \Sigma^{\rho\sigma }_{+\alpha +\beta}   & \, \,\rightarrow \, \,\left( \sigma^{\rho\sigma }\right) _{a}^{\,\, b}, \quad   \Sigma^{\rho\sigma }_{-\alpha -\beta}   \,\, \rightarrow\, \,\left( \bar{\sigma}^{\rho\sigma } \right) ^{\dot{a}}_{\,\,\dot{b}}\ .
    \label{7-03}
\end{IEEEeqnarray}
Considering the Majorana superfields of the Section \ref{sec:RealityCon}, we write
\begin{IEEEeqnarray}{rl}
            \Psi^{+}   & \, = \, 2\sqrt{2}m_{_{\text{D}}}\begin{pmatrix}
\varphi_{a}\\ 
\bar{\psi}^{\dot{a}}
\end{pmatrix} , \quad   \bar{\Psi}^{-}  \,= \, 2\sqrt{2}m_{_{\text{D}}}\begin{pmatrix}
{\psi}^{{a}} & 
\bar{\varphi}_{\dot{a}}
\end{pmatrix} , \nonumber \\
     \Psi^{-}  & \, = \, 2\sqrt{2 } m_{_{\text{D}}} \begin{pmatrix}
\psi_{a}\\ 
\bar{\varphi}^{\dot{a}}
\end{pmatrix} , \quad   \bar{\Psi}^{+}  \, = \, 2\sqrt{2}m_{_{\text{D}}} \begin{pmatrix}
{\varphi}^{{a}} & 
\bar{\psi}_{\dot{a}}
\end{pmatrix} ,
    \label{7-04}
\end{IEEEeqnarray}
where the above two-component superfields satisfy the chiral conditions (see Eq. \eqref{SUSYchiralCondition}):
\begin{IEEEeqnarray}{rl}
             \bar{D}^{\dot{b}}\,\varphi_{a}  \, = \,   \bar{D}^{\dot{b}}\,\bar{\psi}^{\dot{a}}  \, = \, D_{b}\,\psi_{a}   \, = \,  {D}_{b}\,\bar{\varphi}^{\dot{a}}  \, = \,  0\ .
     \label{7-05}
 \end{IEEEeqnarray} 
 These two-spinor superfields, can be expanded in components as
\begin{IEEEeqnarray}{rl}
          \varphi_{a} & \, =\,  \eta_{a}    \, + \,   \theta_{a} \,h\, + \, \theta^{b}f_{ab}\, + \, \theta \theta\,\left(  \tau_{a}  \,- \, i \left( \sigma^{\mu}\right)_{a\dot{b}} \partial_{\mu}\bar{\zeta}^{\,\dot{b}}  \right) \ , 
\label{7-06} \\
          \bar{\psi}^{\dot{a}} & \, =\,\bar{\zeta} ^{\,\dot{a}}    \, + \, {\theta}_{b} \, g^{\dot{a}b}   \, + \, \theta \theta\,\left(\bar{\kappa}^{\dot{a}}    \,- \, i \left( \bar{\sigma}_{\mu}\right) ^{\dot{a}b}\partial^{\mu}{\eta}_{b} \right) \ ,      
     \label{7-07}
 \end{IEEEeqnarray}  
 with $ f_{ab} =f_{ba} $. The  supersymmetric action \eqref{FreeSuperDiracAction} for the Majorana case, written in terms of these superfields, is
\begin{IEEEeqnarray}{rl} 
 A^{\text{Majorana}}_{0}  &  \, = \,     -({1}/4)\int d^{4}x\,d^{2}\theta\, d^{2}\bar{\theta}\,\left[\psi^{a}\varphi_{a}  \, + \, \bar{\varphi}_{\dot{a}}\bar{\psi}^{\dot{a}}\right]     \nonumber \\
 & \qquad \, -\, \int d^{4}x\,d^{2}\theta  \,\left[\tfrac{1}{2}\,\varphi^{a}    \overleftrightarrow{\partial}_{a\dot{b}}\,\bar{\psi}^{\dot{b}} \, + \,  m_{_{\text{D}}}\left( \varphi^{a}\varphi_{a}  \, + \, \bar{\psi}_{\dot{a}}\bar{\psi}^{\dot{a}} \right)\right] \, +\,  \text{h.c.} \nonumber \\
    \label{7-08}
\end{IEEEeqnarray} 
where  $  \overleftrightarrow{\partial}_{a\dot{b}}  \, = \, -i\left(\overrightarrow{\partial}^{\mu}  \, + \, \overleftarrow{\partial}^{\mu} \right) \left(\sigma_{\mu}\right)_{a\dot{b}}  $. The relation between the component fields  ($ \chi^{+}_{\alpha} $, $ \mathcal{G}^{+}_{\alpha}  $, $ d $, $ \phi $, $ a_{\mu} $,  $ b_{\mu} $, $ c_{\mu\nu} $) and the  set ($\eta_{a} $, $\bar{\zeta}^{\dot{a}} $, $ h $, $ g^{a\dot{b}} , f_{ab} $), carrying Dirac-Lorentz and dotted-undotted indices, respectively, is given by
\begin{IEEEeqnarray}{rl}
            \chi_{\alpha}^{+}  \, = \, 2\sqrt{2}m_{_{\text{D}}}\begin{pmatrix}
\eta_{a} \\ 
\bar{\zeta}^{\dot{a}} 
\end{pmatrix} , \quad
            \mathcal{G}_{\alpha}^{+}  \, = \,2\sqrt{2}m_{_{\text{D}}} \begin{pmatrix}
\tau_{a} \\ 
\bar{\kappa}^{\dot{a}} 
\end{pmatrix} , \quad     
    \label{7-09}
\end{IEEEeqnarray}
and
\begin{IEEEeqnarray}{rl}
     h   &\, = \, 2\sqrt{2}m_{_{\text{D}}}\left(  id  \, - \, m_{_{\text{D}}}\phi\right)  \ , \nonumber \\
    g^{\dot{a}b}  & \, = \, 2\sqrt{2} m_{_{\text{D}}}\left(\bar{\sigma}_{\mu}\right)^{\dot{a}b}\left(  m_{_{\text{D}}} a^{\mu} \, - \, i b^{\mu}  \, - \,i\partial^{\mu}\phi\right) \ , \nonumber \\
           f_{ab}  & \, = \,-2\sqrt{2}m_{_{\text{D}}}\left( \sigma^{\mu\nu}\right)_{ab} \left[ f_{\mu\nu}  \, + \, c_{\mu\nu} \right] \, \quad \ ,
    \label{7-10}
\end{IEEEeqnarray}
with $ f^{\mu\nu}  \, = \,   \partial ^{\mu}a^{\nu}\, - \,  \partial ^{\nu}a^{\mu} $.
All fields with Lorentz indices (including scalars) are real [see Eq. \eqref{MajoranaBosonSector}] and the remaining fields  $ \chi^{-} $ and   $ \mathcal{G}^{-} $  are defined according to Eq. \eqref{MajoranaFermionSector}.
 
 To obtain the massive  one-half superspin model  of Ref. \cite{Siegel:1979ai}, we only consider one chiral superfield:
  \begin{IEEEeqnarray}{rl}
        \phi_{a} & \, =\,  \eta_{a}    \, + \,   \theta_{a} \left(A  \, + \, i B \right) \, + \, \theta^{b}F_{ab}\, + \, \theta \theta\,\left(  \chi_{a}  \,- \, i \left( \sigma^{\mu}\right)_{a\dot{b}} \partial_{\mu}\bar{\eta}^{\,\dot{b}}  \right) \ , 
\nonumber\\          
     \label{7-11}
 \end{IEEEeqnarray}  
 with $ A={A}^{*} $, $ B=B^{*} $ and $ F_{ab}=F_{ba} $.  The propagating fields are $  \eta_{a} , A, F_{ab}$, while $ B $ is an auxiliary superfield. They do not satisfy any additional constraints. The spin 1 boson is encoded in the real second-rank  antisymmetric  field $  \tilde{B}_{\mu\nu}  $,
\begin{IEEEeqnarray}{rl}
            F_{ab} & \, = \,-(1/2\sqrt{2})\left( \sigma^{\mu\nu}\right)_{ab}   \tilde{B}_{\mu\nu}\ .
    \label{7-12}
\end{IEEEeqnarray}
Since this tensor field  does not satisfy  the Bianchi equations, it cannot be decomposed as derivatives of a vector field [the  free action for this field  is given by Eq. \eqref{2AntiSymTensor-MassiveAction}]. 
The supersymmetric action  \eqref{Siegel-02}, in a two-spinor notation (for the case of the neutral superparticle) acquires the form
\begin{IEEEeqnarray}{rl}
        A_{\text{Siegel}}   \, = \,   -\tfrac{1}{4}\int d^{4}x d^{2}\theta d^{2}\bar{\theta}\left(  D^{a}\phi_{a}  \, + \, \bar{D}^{\dot{a}}\bar{\phi}_{\dot{a}}\right)^{2}  \,   \, - \, \left( { m^{2}_{_{\text{D}}}}/{2}\right) \int d^{4}x \,d^{2}\theta\,\phi^{a}\phi_{a}   \, + \, \text{h.c.} \nonumber \\
    \label{7-13}
\end{IEEEeqnarray}
or alternatively:
\begin{IEEEeqnarray}{rl}
  A_{\text{Siegel}}    \, = \,           -\tfrac{1}{4}\int d^{4}x d^{2}\theta d^{2}\bar{\theta}\left[  \,\phi^{a}    \overleftrightarrow{\partial}_{a\dot{b}}\,\bar{\phi}^{\dot{b}} \right]  \, + \, \tfrac{1}{4}\int d^{4}x d^{2}\theta\left[ \phi^{a} \left(\square  \, - \,2 m^{2}_{_{\text{D}}}\right)  \phi_{a} \right] \, + \, \text{h.c.} \nonumber \\
\label{7-14}
\end{IEEEeqnarray}
One of the main differences  between the actions  \eqref{7-08}
and \eqref{7-14}, is that the introduction of an extra superfield $   \bar{\psi}^{\dot{a}} $  allows to dispense of an extra derivative operator in the bilinear terms of the supersymmetric Lagrangian. Note that the  first term of   the superpotential in  \eqref{7-08},  mix the superfields $ \varphi^{a}$ and  $ \psi^{a}$, making impossible to split  \eqref{7-08} into separate actions for each superfield. It cannot be stressed too strongly, that the differences between the model of Siegel and ours, will only appear in an interacting theory, as exemplified in Section \ref{sec:SupYuk}. 
\section{Conclusions}
\label{sec:Conclusions}
In this paper we have constructed the theory for  massive Dirac chiral superfields by eliminating the second order derivatives in fermionic fields from the free Lagrangian. The strategy employed, in order to identify the propagating and auxiliary bosonic fields, was to look for the most general  tensor-spinor field in the interaction picture and then promote it to the off-shell case.

The  Dirac superfields are only constrained by the (supersymmetric) chiral conditions, in contrast to the case of a massless superhelicity one-half supermultiplet, where strength superfields satisfy supersymmetric Bianchi identities instead.  

We also show that the mass parameter appears explicitly in the bosonic propagating fields; this feature is not exclusive of this formulation~\cite{Siegel:1979ai}.  We point out that our formulation differs crucially from the known models for superspin one-half: in those models the elimination of the auxiliary fields restores the free Lagrangian of the bosonic sector, whereas in our case it does it for fermions. In this sense, our model resembles the Wess-Zumino model, as in both models the zeroth and second order  components  of the chiral superfields are the propagating and auxiliary fields, respectively. The similarity however is not present in the linear component.

 A key ingredient of our formulation is the introduction of Dirac derivative terms in the $ \mathcal{F} $-terms of the superfields. Without them, the Dirac kinetic term would only appear after the elimination of the auxiliary fields, thus leading to second order fermions off-shell, an undesired feature.

Our result  is very simple in the sense that we have added to the Lagrangian one more supersymmetric bilinear with respect to the free Wess-Zumino model. Besides, we have shown that the free theory of our model is equivalent to the current known models of massive one-half superspin. Indeed, the free bare superpropagators of Siegel model and ours coincide. Interestingly, the presence of only one superderivative operator in our framework can potentially lead to a different behavior with respect to Siegel model (where two superderivatives are present) when supergravity effects are taken into account (this case  is currently under investigation).    

The formalism presented includes the case where the Dirac superfield satisfies the Majorana condition, reducing its dynamical components to two Majorana fields, a real scalar, and  a real vector. An interaction theory  with a super Yukawa coupling has been presented in terms of the off-shell component fields.  We have shown that this coupling can not be present in Siegel model.  Further study on the renormalization group equations for the couplings  of the model is required. In particular it would be interesting to determine if  this class of models can provide a scenario, besides that of spontaneously broken gauge symmetries,  for renormalizable theories with massive vector fields.

Another attractive direction for future study consists on applying the present formalism to more general cases with arbitrary odd-superspin. An interesting example  is the superspin three-halves with Rarita-Schwinger chiral superfields, in which the bosonic sector contains massive vectors and gravitons. 

Finally, for the superspin one-half case, a full equivalence between the canonical and path integral methods of this work and the noncanonical formalism presented in \cite{Jimenez:2014gfa} has been established from the structure of the superpropagators and the explicit form of the interaction picture superfields.

\section*{Acknowledgments}
 We  would like to thank A. Aranda  for useful discussions and for corrections to the manuscript and also to the University  of Colima for their hospitality during the completion of this work. E.J. acknowledges support from the Mexican grants: PAPIIT IN113712 and CONACyT-132059. C.A.V-A. acknowledges the Spanish grants FPA2014-58183-P, Multidark CSD2009-00064, SEV-2014-0398 (MINECO), PROMETEOII/2014/084 (Generalitat Valenciana) and support from CONACyT (Mexico) under grant 251357.
\appendix
\section{Notation and conventions}\label{Sec:ConventionsNotation}
In this appendix we provide formulae for the spinor-tensor fields that are used throughout the paper, we include  the derivation of relation \eqref{ExpansionPropagatingFields} for  the  tensor-spinor field in the interaction picture. First we establish our notation and conventions. 
 
Dirac indices are labeled  by $ \alpha,\alpha',\beta,\beta' $, etc., Lorentz indices by   $ \mu,\nu,\mu',\nu' $, etc. We take the Lorentz metric as $ \eta_{\mu\nu}  \, = \, \text{diag}\left(1,1,1,-1 \right)   $ and the anticommutator of   $ \gamma $-matrices is given by  $  \left\lbrace  \gamma^{\mu} ,  \gamma^{\nu}\right\rbrace   \, = \, 2\eta^{\mu\nu} $ .  We stick to the representation
\begin{equation}
         \gamma^{0}\, = \,-i\begin{pmatrix}
0 & I \\ 
I & 0
\end{pmatrix}  \, = \, -i\beta, \quad {\gamma}_{i}\, = \,-i  \begin{pmatrix}
0 & \sigma_{i} \\ 
-\sigma_{i} & 0
\end{pmatrix}, \quad 
     \gamma_{5}  \, = \, \begin{pmatrix}
I & 0 \\ 
0 & -I
\end{pmatrix} \ ,         \label{Appendix_GammaWeylBasis}
	\end{equation}	
\begin{IEEEeqnarray}{rl}
 \epsilon  \, = \, \begin{pmatrix}
 e & 0 \\ 
 0 & e
 \end{pmatrix} , \quad e  \, = \, \begin{pmatrix}
 0 & 1 \\ 
 -1 & 0
 \end{pmatrix} \ .
    \label{Appendix_Transposing}
\end{IEEEeqnarray}
The matrices $ \beta $ and $ \epsilon\gamma_{5} $ satisfy
 \begin{IEEEeqnarray}{rl}
               \beta \gamma^{\mu}   \, = \,  -\gamma^{\mu\dagger}\beta, \quad   \epsilon\gamma_{5} \gamma^{\mu} \, = \, -\gamma^{\mu \intercal}\epsilon\gamma_{5}\ .
     \label{Appendix_Transpos_ConjugateOfDiracMatrices}
 \end{IEEEeqnarray}	
 These are the same conventions of  \cite{Weinberg:2000cr}, except for left and right spinors, which we  label  with subindices plus and minus, respectively.
 
We expand the tensor-spinor $ \xi_{\alpha\beta} $ in the  the $4\times 4$  covariant basis $ \left(I,\gamma_{5},\gamma_{\mu},\gamma_{\mu}\gamma_{5},\left[\gamma_{\mu},\gamma_{\nu} \right]  \right)  $ as
\begin{IEEEeqnarray}{rl}
         \sqrt{2} \xi &\, = \,\left(   \xi_{0}\, + \, \xi_{5}\gamma_{5}  \, + \, \xi_{\rho} \gamma^{\rho}  \, + \,      \xi_{5\rho}\gamma^{\rho}\gamma_{5}  \, + \,  \tfrac{1}{4}\xi_{\rho\sigma} \left[ \gamma^{\rho} ,\gamma^{\rho} \right]\right)  \left[ \epsilon\gamma_{5}\right]  \ .     \label{Appendix_TensorSpinorCovBasis}
\end{IEEEeqnarray}
The Dirac adjoint of $ {\xi} $ is
\begin{IEEEeqnarray}{rl}          
            \bar{\xi}  & \, \equiv \, \beta \xi^{*} \beta \ ,
    \label{ElAsunto}
\end{IEEEeqnarray}
and its expansion in the invariant basis reads 
\begin{IEEEeqnarray}{rl}
         \sqrt{2} \,\bar{\xi}  &\, = \, \left[- \epsilon\gamma_{5}\right] \left(  \xi^{*}_{0}  \, - \, \xi^{*}_{5}\gamma_{5}  \, + \, \xi^{*}_{\mu} \gamma^{\mu}  \, + \,      \xi^{*}_{5\mu}\gamma^{\mu}\gamma_{5}  \, + \, \tfrac{1}{4}\xi^{*}_{\mu\nu} \left[ \gamma^{\mu} ,\gamma^{\nu} \right] \right) \ .  
    \label{ComoTeVesMe1}
\end{IEEEeqnarray}

Let $ \bar{\xi} $ and $ \xi' $ be two tensor-spinors, we adopt the following notation for the the trace of their product:
\begin{IEEEeqnarray}{rl}
             \bar{\xi}\xi'  \, \equiv \, \sum_{\alpha\beta}\bar{\xi}_{\alpha\beta}\,\xi'_{\alpha\beta} \  .
    \label{DiracBilinears}
\end{IEEEeqnarray}
In terms of the  symmetric  and antisymmetric parts of the tensor-spinor fields
\begin{IEEEeqnarray}{rl}
            \xi  \, = \, \xi_{a}  \, + \, \xi_{s}, \quad  \xi_{a}  \, = \, -\left(  \xi_{a}\right)^{\intercal}, \quad    \xi_{s}  \, = \, \left( \xi_{s}\right)^{\intercal} \ ,
    \label{EsoEsTodo2}
\end{IEEEeqnarray}
we can write
 \begin{IEEEeqnarray}{rl}
         \sqrt{2} \,\xi_{a} &\, = \,\left(   \xi_{0}\, + \, \xi_{5}\gamma_{5}  \, + \,      \xi_{5\rho}\gamma^{\rho}\gamma_{5}\right)  \left[ \epsilon\gamma_{5}\right],  \nonumber \\
          \sqrt{2}\, \xi_{s} &\, = \,\left( \xi_{\rho} \gamma^{\rho} \, + \, \tfrac{1}{4}\xi_{\rho\sigma} \left[ \gamma^{\rho} ,\gamma^{\sigma} \right]\right)  \left[ \epsilon\gamma_{5}\right],   \nonumber \\
               \sqrt{2} \,\bar{\xi}_{a} &\, = \, \left[- \epsilon\gamma_{5}\right] \left(  \xi^{*}_{0}  \, - \, \xi^{*}_{5}\gamma_{5}   \, - \,      \xi^{*}_{5\mu}\gamma^{\mu}\gamma_{5}  \right),   \nonumber \\ 
          \sqrt{2}\, \bar{\xi}_{s} &\, = \, \left[- \epsilon\gamma_{5}\right] \left[   \xi^{*}_{\mu} \gamma^{\mu}   \, + \, \tfrac{1}{4}\xi^{*}_{\mu\nu} \left[ \gamma^{\mu} ,\gamma^{\nu} \right] \right)  \ .
    \label{ComoTeVesMe2}
\end{IEEEeqnarray}
Notice that $ \bar{\xi}_{a,s} \, = \, \overline{\left( {\xi}_{a,s}\right) } $. For any $ 4\times 4 $ matrix $ M_{\alpha\beta} $, the relation
\begin{IEEEeqnarray}{rl}
             \tfrac{1}{2}  \bar{\xi}\left( M\,\xi'  \, + \, \xi'\,M^{\intercal}\right)   \, = \,           \bar{\xi}_{a} M\,\xi'_{a}     \, + \,         \bar{\xi}_{s} M\,\xi'_{s}   \  
             \label{SymmetricContractionSpinor}
\end{IEEEeqnarray}
holds. In the particular cases $M = I, \slashed{\partial} $; each term on the right side of \eqref{SymmetricContractionSpinor} gives
\begin{IEEEeqnarray}{rl}
            -\tfrac{1}{2} \bar{\xi}_{a} {\xi}'_{a}   & \, = \, \xi^{*}_{0} \xi'_{0}   \, - \, \xi^{*}_{5} \xi'_{5}   \, + \,\xi^{*}_{5\mu}\xi_{5}'^{\mu},   \nonumber \\
                 - \tfrac{1}{2} \bar{\xi}_{s} {\xi}'_{s}  & \, = \,  - \xi^{*}_{\mu}\xi'^{\mu}  \, + \,\tfrac{1}{2} \xi^{*}_{\mu\nu} \xi'^{\mu\nu},     \nonumber \\
             -\tfrac{1}{2} \bar{\xi}_{a}\slashed{\partial}   {\xi}'_{a}   & \, = \,  \xi^{*}_{5}\partial_{\mu}\xi'^{\mu}_{5} \, -\, \xi^{\mu *}_{5}\partial_{\mu}\xi'_{5},    \nonumber \\              
                - \tfrac{1}{2} \bar{\xi}_{s}\slashed{\partial}  {\xi}'_{s}  & \, = \,     \xi^{*}_{\nu}\partial_{\mu} \xi'^{\mu\nu}\, + \,\xi^{\mu\nu *}\partial_{\mu}\xi'_{\nu}  \  .
    \label{QuePedo4}
\end{IEEEeqnarray}
When  $  \xi_{\alpha\beta} $ is of the form  $ \psi_{\alpha}\chi_{\beta} $ we have
\begin{IEEEeqnarray}{rl}
         {2\sqrt{2}} \,  \xi_{0}   & \, = \,  \psi^{\intercal}\epsilon\gamma_{5} \chi , \quad 
        {2\sqrt{2}} \,    \xi_{5}      \, = \,  \psi^{\intercal}\epsilon \chi , \quad \nonumber \\   
         {2\sqrt{2}} \,    \xi_{\mu} &   \, = \, - \psi^{\intercal}\epsilon\gamma_{5}\gamma_{\mu} \chi ,\quad 
             {2\sqrt{2}} \,    \xi_{5\mu}  \, = \,  \psi^{\intercal}\epsilon\gamma_{\mu} \chi, \nonumber \\
              -{16\sqrt{2}} \,    \xi_{\mu\nu} &   \, = \, - \psi^{\intercal}\epsilon\gamma_{5}\left[ \gamma_{\mu},\gamma_{\nu} \right] \chi \ .
\label{Appendix_CovariantBasisDiracTensor}
\end{IEEEeqnarray}

The rest of this Appendix focuses on  the decomposition of the free tensor-spinor field operator  in terms of   the scalar and vector free fields.   The  free causal tensor-spinor  field has the form
\begin{IEEEeqnarray}{rl}
              \xi_{\alpha\beta} ^{\text{free.}}   &\, = \,     {(2\pi)^{-\frac{3}{2}}}  \,\sum^{}_{\sigma,{\sigma}' } \int d^{3}\textbf{p} \left( {2p^{0}}\right) ^{\frac{1}{2}}\, \left\lbrace \,e^{ +i  x\cdot p  }\, b(\mathbf{p},\sigma;\sigma') {u}_{\alpha}(\textbf{p} ,\sigma)  u_{\beta}(\textbf{p} ,{\sigma}')    \right.  \nonumber \\
  &  \left.         \, + \,   \, e^{ -ix\cdot p } \, \left[ b^{c}(\mathbf{p},\sigma;{\sigma}') \right] ^{*}{v}_{\alpha }(\textbf{p} ,\sigma)   v_{\beta}(\textbf{p} ,{\sigma}')  \right\rbrace   \ .    \label{Appendix_FreeTensorSpinorField}
\end{IEEEeqnarray}
The annihilation  particle operator   $ b(\mathbf{p},\sigma;{\sigma}')  $ and  the creation antiparticle operator   $ b^{c *}(\mathbf{p},\sigma;{\tilde{\sigma}})  $  carry the projection $ (\sigma,{\tilde{\sigma}} )$ of spin $ \frac{1}{2}\otimes\frac{1}{2}  $. The spinors $ {u}_{\alpha} $ and $ v_{\alpha} $, are the usual Dirac wavefunctions.  Let   $ b^{0} $  and $ b^{1} $  be the annihilation operators  of definite spin zero and one, respectively.  Thus 
\begin{IEEEeqnarray}{rl}
            b(\mathbf{p},\sigma; {\sigma}')   &  \, = \,     \sum^{+1}_{\tilde{\sigma}= -1 }C_{\frac{1}{2},\frac{1}{2}}\left( 1,\,\tilde{\sigma}\, ; \sigma,\sigma'\right)   b^{1}\left( \mathbf{p},\tilde{\sigma}\right)     \, + \, C_{\frac{1}{2},\frac{1}{2}}\left(0,0\, ; \sigma,\sigma'\right)   b^{0}\left( \mathbf{p}\right)   \ , 
\label{Appendix_AsDefiniteAngularMomentum}
\end{IEEEeqnarray}
where the constants in the right-hand side are the Clebsch-Gordan coefficients for the decomposition of the state with angular momentum $ \frac{1}{2}\otimes \frac{1}{2} $ into states of  definite spin. Inserting \eqref{Appendix_AsDefiniteAngularMomentum} in  \eqref{Appendix_FreeTensorSpinorField}  
we obtain the wavefunctions  for the fields that carry zero and one spins. For the former case we have
\begin{IEEEeqnarray}{rl}
             u^{0}_{\alpha\beta}(\textbf{p})    &\, \equiv \, \sqrt{2p^{0}}\sum^{+\frac{1}{2}}_{\sigma = -\frac{1}{2} } \,\sum^{\frac{1}{2}}_{{\sigma}'  =-\frac{1}{2} }     C_{\frac{1}{2},\frac{1}{2}}\left(0,\,0\, ; \sigma,\sigma'\right)    u_{\alpha}(\textbf{p} ,{\sigma}')  u_{\beta}(\textbf{p} ,{\sigma})\ , \nonumber \\
               v^{0}_{\alpha,\beta}(\textbf{p})    &\, \equiv \, \sqrt{2p^{0}}\sum^{+\frac{1}{2}}_{\sigma = -\frac{1}{2} } \,\sum^{\frac{1}{2}}_{{\sigma}'  =-\frac{1}{2} }     C_{\frac{1}{2},\frac{1}{2}}\left(0,\,0\, ; \sigma,\sigma'\right)    v_{\alpha}(\textbf{p} ,{\sigma}') v_{\beta}(\textbf{p} ,\sigma)\ . 
     \label{WaveFunctions}
 \end{IEEEeqnarray} 
The only non-vanishing Clebsch-Gordan coefficients are
\begin{IEEEeqnarray}{rl}
            C_{\frac{1}{2},\frac{1}{2}}\left(0 \, ;\tfrac{1}{2},-\tfrac{1}{2} \right)   \, = \,    \left( - \right)   C_{\frac{1}{2},\frac{1}{2}}\left(0 \, ; -\tfrac{1}{2},\tfrac{1}{2}\right)   \, = \,   1/\sqrt{2} \ .     \label{NonVanishing}
\end{IEEEeqnarray}
With the help of \eqref{Appendix_CovariantBasisDiracTensor} we write  $  u^{0}_{\alpha,\beta}(\textbf{p})  $  in the covariant basis \eqref{Appendix_TensorSpinorCovBasis} as
\begin{IEEEeqnarray}{rl}
             u^{0}_{\alpha\beta}(\textbf{p},\sigma)  \, = \,  \frac{1}{\sqrt{2}\sqrt{2p^{0}}}\left[ \left(m  \, + \,{i\slashed{p}} \right) \epsilon\right] _{\alpha\beta}\ .
    \label{AntiSymme1}
\end{IEEEeqnarray}
Similarly, for the creation part we have
\begin{IEEEeqnarray}{rl}
             v^{0}_{\alpha\beta}(\textbf{p} )  \, = \,  \frac{1}{\sqrt{2}\sqrt{2p^{0}}}\left[ \left(m  \, - \,{i\slashed{p}} \right) \epsilon \right] _{\alpha\beta}\ .
    \label{AntiSymme2}
\end{IEEEeqnarray}
 
For the spin one sector we obtain the following decomposition:
\begin{IEEEeqnarray}{rl}
         u^{1}_{\alpha\beta}(\textbf{p} ,{\sigma})    &   \, = \,\frac{- 1}{\sqrt{2} \sqrt{2p^{0}}} \left[ i\,m\, e_{\mu }\left(p, \sigma \right)  \, + \,\frac{1}{2}\left(p_{\nu}  e_{\mu }\left(p, \sigma \right)  \, - \, p_{\mu} e_{\nu }\left(p, \sigma \right)\right)  \gamma^{\nu} \right]\gamma^{\mu}\epsilon\gamma_{5}   \ ,\nonumber \\
           v^{1}_{\alpha,\beta}(\textbf{p} ,{\sigma})    &   \, = \,\frac{- 1}{\sqrt{2} \sqrt{2p^{0}}} \left[ i\,m\, e^{*}_{\mu }\left(p, \sigma \right)  \, - \,\frac{1}{2}\left( p_{\nu} e^{*}_{\mu }\left(p, \sigma \right)  \, - \, p_{\mu} e^{*}_{\nu }\left(p, \sigma \right)\right)  \gamma^{\nu} \right]\gamma^{\mu}\epsilon\gamma_{5} \ ,
    \label{AInDiracBasis}
\end{IEEEeqnarray} 
where $ \left( {\sigma} =-1,0,+1 \right) $ and $  e_{\mu }\left(p, \sigma \right) $  are the  wavefunctions for massive vector fields. Finally, defining the  scalar and vector free fields as
\begin{IEEEeqnarray}{rl}                  
  \phi^{\text{free}}   &  \, = \,      (2\pi)^{-\frac{3}{2}}  \int \frac{d^{3}\textbf{p}}{\sqrt{2 p^{0}}}\, \left\lbrace \,e^{ +i  x\cdot p  }\, b^{0 }(\mathbf{p})       \, + \,    e^{ -ix\cdot p } \, \left[  b^{0, c }(\mathbf{p})     \right] ^{*}  \right\rbrace   \ ,\nonumber \\
  a_{\mu}^{\text{free}}   &  \, = \,    (2\pi)^{-\frac{3}{2}}\sum^{+1}_{\sigma=-1} \int \frac{d^{3}\textbf{p}}{\sqrt{2 p^{0}}}\, \left\lbrace \,e^{ +i  x\cdot p  }\,e_{\mu} \left( \mathbf{p}, \sigma\right) b^{1 }(\mathbf{p},\sigma)       \, + \,    e^{ -ix\cdot p } \,e^{*}_{\mu } \left( \mathbf{p}, \sigma\right) \left[  b^{1,c }(\mathbf{p},\sigma)     \right] ^{*}  \right\rbrace   \ , \nonumber \\
    \label{FreeBosonicFields}
\end{IEEEeqnarray}
we arrive at \eqref{ExpansionPropagatingFields}.

\bibliographystyle{elsarticle-harv}

\bibliography{Dirac_Superfields_Jimenez}{}

\begin{thebibliography}{10}

\bibitem{Wess:1992cp}
J.~Wess and J.~Bagger.
\newblock {\em {Supersymmetry and supergravity}}.
\newblock 1992.

\bibitem{Buchbinder:1995uq}
I.~L. Buchbinder and S.~M. Kuzenko.
\newblock {\em {Ideas and methods of supersymmetry and supergravity: A Walk
  through superspace}}.
\newblock 1995.

\bibitem{Ogievetsky:1976qb}
V.~I. Ogievetsky and E.~Sokatchev.
\newblock {Superfield Equations of Motion}.
\newblock {\em J. Phys.}, A10:2021--2030, 1977.

\bibitem{Altendorfer:1999mn}
Richard Altendorfer and Jonathan Bagger.
\newblock {Dual supersymmetry algebras from partial supersymmetry breaking}.
\newblock {\em Phys. Lett.}, B460:127--134, 1999.

\bibitem{Buchbinder:2002gh}
I.~L. Buchbinder, S.~James Gates, Jr., William~Divine Linch, III, and
  J.~Phillips.
\newblock {New 4-D, N=1 superfield theory: Model of free massive superspin 3/2
  multiplet}.
\newblock {\em Phys. Lett.}, B535:280--288, 2002.

\bibitem{Buchbinder:2002tt}
I.~L. Buchbinder, S.~James Gates, Jr., William~Divine Linch, III, and
  J.~Phillips.
\newblock {Dynamical superfield theory of free massive superspin-1 multiplet}.
\newblock {\em Phys. Lett.}, B549:229--236, 2002.

\bibitem{Gregoire:2004ic}
Thomas Gregoire, Matthew~D. Schwartz, and Yael Shadmi.
\newblock {Massive supergravity and deconstruction}.
\newblock {\em JHEP}, 07:029, 2004.

\bibitem{Buchbinder:2005je}
I.~L. Buchbinder, S.~James~Gates, Jr., S.~M. Kuzenko, and J.~Phillips.
\newblock {Massive 4D, N=1 superspin 1 \& 3/2 multiplets and dualities}.
\newblock {\em JHEP}, 02:056, 2005.

\bibitem{Gates:2005su}
S.~James Gates, Jr. and Sergei~M. Kuzenko.
\newblock {4D, N = 1 higher spin gauge superfields and quantized twistors}.
\newblock {\em JHEP}, 10:008, 2005.

\bibitem{Gates:2006cq}
S.~James Gates, Jr., Sergei~M. Kuzenko, and Gabriele Tartaglino-Mazzucchelli.
\newblock {New massive supergravity multiplets}.
\newblock {\em JHEP}, 02:052, 2007.

\bibitem{Gates:2013tka}
S.~James. Gates, Jr. and Konstantinos Koutrolikos.
\newblock {A dynamical theory for linearized massive superspin 3/2}.
\newblock {\em JHEP}, 03:030, 2014.

\bibitem{D'Auria:2004sy}
Riccardo D'Auria and Sergio Ferrara.
\newblock {Dyonic masses from conformal field strengths in D even dimensions}.
\newblock {\em Phys. Lett.}, B606:211--217, 2005.

\bibitem{Louis:2004xi}
Jan Louis and Waldemar Schulgin.
\newblock {Massive tensor multiplets in N=1 supersymmetry}.
\newblock {\em Fortsch. Phys.}, 53:235--245, 2005.

\bibitem{Theis:2004pa}
Ulrich Theis.
\newblock {Masses and dualities in extended Freedman-Townsend models}.
\newblock {\em Phys. Lett.}, B609:402--407, 2005.

\bibitem{Kuzenko:2004tn}
Sergei~M. Kuzenko.
\newblock {On massive tensor multiplets}.
\newblock {\em JHEP}, 01:041, 2005.

\bibitem{Siegel:1979ai}
Warren Siegel.
\newblock {Gauge Spinor Superfield as a Scalar Multiplet}.
\newblock {\em Phys. Lett.}, B85:333, 1979.

\bibitem{Gates:1980az}
S.~James Gates, Jr. and W.~Siegel.
\newblock {VARIANT SUPERFIELD REPRESENTATIONS}.
\newblock {\em Nucl. Phys.}, B187:389, 1981.

\bibitem{Weinberg:1969di}
Steven Weinberg.
\newblock {Feynman rules for any spin. iii}.
\newblock {\em Phys. Rev.}, 181:1893--1899, 1969.

\bibitem{Jimenez:2014gfa}
Enrique Jiménez.
\newblock {$\mathcal N=1$ super Feynman rules for any superspin: Noncanonical
  SUSY}.
\newblock {\em Phys. Rev.}, D92(8):085013, 2015.

\bibitem{Gates:2001rn}
S.~James Gates, Jr., William~Divine Linch, III, J.~Phillips, and L.~Rana.
\newblock {The Fundamental supersymmetry challenge remains}.
\newblock {\em Grav. Cosmol.}, 8:96--100, 2002.

\bibitem{Dixon:2015cya}
John~A. Dixon.
\newblock {An Irreducible Massive Superspin One Half Action Built From the
  Chiral Dotted Spinor Superfield}.
\newblock {\em Phys. Lett.}, B744:244--249, 2015.

\bibitem{Wunderle:2010yw}
Kai~E. Wunderle and Rainer Dick.
\newblock {A Supersymmetric Lagrangian for Fermionic Fields with Mass Dimension
  One}.
\newblock {\em Can. J. Phys.}, 90:1185--1199, 2012.

\bibitem{Veltman:1997am}
M.~J.~G. Veltman.
\newblock {Two component theory and electron magnetic moment}.
\newblock {\em Acta Phys. Polon.}, B29:783--798, 1998.

\bibitem{Weinberg:2000cr}
Steven Weinberg.
\newblock {\em {The quantum theory of fields. Vol. 3: Supersymmetry}}.
\newblock Cambridge University Press, 2013.

\bibitem{Kalb:1974yc}
Michael Kalb and Pierre Ramond.
\newblock {Classical direct interstring action}.
\newblock {\em Phys. Rev.}, D9:2273--2284, 1974.

\bibitem{VanProeyen:1979ks}
Antoine Van~Proeyen.
\newblock {Massive Vector Multiplets in Supergravity}.
\newblock {\em Nucl. Phys.}, B162:376, 1980.

\bibitem{Mukhi:1979wc}
Sunil Mukhi.
\newblock {MASSIVE VECTOR MULTIPLET COUPLED TO SUPERGRAVITY}.
\newblock {\em Phys. Rev.}, D20:1839, 1979.

\end{thebibliography}

\end{document}